\documentclass[aps,prc,twocolumn,nofootinbib,superscriptaddress,longbibliography]{revtex4-2}
\usepackage{amsfonts}
\usepackage{amscd}
\usepackage{amsmath}
\usepackage{amssymb}
\usepackage{newtxtext,newtxmath}
\usepackage{here}
\usepackage[dvipdfmx]{graphicx}
\usepackage{dcolumn}
\usepackage{bm}
\usepackage{xcolor}
\usepackage[%
  colorlinks=true,
  urlcolor=blue,
  linkcolor=blue,
  citecolor=blue
]{hyperref}
\usepackage{tabularx}
\usepackage{etoolbox}
\usepackage{braket}
\usepackage[caption=false]{subfig}
\newcommand{\nn}{\nonumber}
\newcommand{\beq}{\begin{equation}}
\newcommand{\eeq}{\end{equation}}

\makeatletter
\let\cat@comma@active\@empty
\makeatother
\renewcommand{\vec}[1]{\boldsymbol{#1}}

\allowdisplaybreaks

\usepackage[normalem]{ulem}

\renewcommand\sout{\bgroup \color{blue} \ULdepth=-.5ex \ULset}

\usepackage{hyperref}

\begin{document}

\title{Spin-polarized phases of $^3P_2$ superfluids in neutron stars}

\author{Takeshi Mizushima}
\email{mizushima@mp.es.osaka-u.ac.jp}
\affiliation{Department of Materials Engineering Science, Osaka University, Toyonaka, Osaka 560-8531, Japan}
\author{Shigehiro Yasui}
\email{yasuis@keio.jp}
\affiliation{Department of Physics \& Research and Education Center for Natural Sciences, Keio University, Hiyoshi 4-1-1,
Yokohama, Kanagawa 223-8521, Japan}
\affiliation{Interdisciplinary Theoretical and Mathematical Sciences Program (iTHEMS), RIKEN, Wako, Saitama 351-0198, Japan}
\author{Daisuke Inotani}
\email{dinotani@keio.jp}
\affiliation{Department of Physics \& Research and Education Center for Natural Sciences, Keio University, Hiyoshi 4-1-1,
Yokohama, Kanagawa 223-8521, Japan}
\author{Muneto Nitta}
\email{nitta@phys-h.keio.ac.jp}
\affiliation{Department of Physics \& Research and Education Center for Natural Sciences, Keio University, Hiyoshi 4-1-1,
Yokohama, Kanagawa 223-8521, Japan}

\date{\today}

\begin{abstract}
The interior of a neutron star is expected to be occupied by a neutron $^3P_2$ superfluid, which is the condensate of spin-triplet $p$-wave Cooper pairs of neutrons with total angular momentum $J=2$. Here we investigate the thermodynamic stability of $^3P_2$ superfluids in a neutron-star interior under a strong magnetic field. Using the theory incorporating the finite size correction of neutron Fermi surface, we show that the spin-polarized phases of $^3P_2$ superfluids, the magnetized biaxial nematic phase and the ferromagnetic phase, appear in high temperatures and high magnetic fields. These phases were missed in the previous studies using the quasiclassical approximation in which dispersions of neutrons are linearized around the Fermi surface. In particular, the ferromagnetic phase, which is the condensation of Cooper-paired neutrons with fully polarized spins, appears between the normal phase and the biaxial nematic phase and enlarge the thermodynamic stability of $^3P_2$ superfluids under strong magnetic fields. Furthermore, we present the augmented Ginzburg-Landau theory that incorporates the thermodynamic stability of spin-polarized $^3P_2$ superfluid phases.
\end{abstract}

\maketitle



\section{Introduction}

%

A neutron star, which is composed almost entirely of dense neutrons, is a colossal superfluid in the Universe~\cite{Chamel2017,Haskell:2017lkl,Sedrakian:2018ydt}. When the temperature of a neutron star falls below a certain critical value it becomes energetically favorable for neutrons to form Cooper pairs. 
The neutron superfluidity is a key ingredient for understanding the vital properties of neutron stars, including the long relaxation time observed in the sudden speed-up events (pulsar glitches) of neutron stars~\cite{Baym1969,Pines1972,Takatsuka:1988kx} and the enhancement of neutrino emission at the onset of superfluid transition~\cite{Yakovlev:2000jp,Potekhin:2015qsa,Yakovlev:1999sk,Heinke:2010cr,Shternin2011,Page:2010aw}. Pulsar glitches may also be explained by the existence of superfluid components with quantized vortices~\cite{REICHLEY1971,Anderson:1975zze}.

The condensates of paired neutrons exist under extreme conditions prevailing in the interiors of neutron stars such as high density, rapid rotation, and a strong magnetic field~\cite{Graber:2016imq,Baym:2017whm}. In particular, magnetars are known as neutron stars accompanied by the most intense magnetic field ($B=10^{15}$-$10^{18}~{\rm G}$) observed in the Universe. Neutrons in the inner crust form spin-singlet $s$-wave Cooper pairs via the $^1S_0$ attractive interaction~\cite{Migdal:1960}, while the $^1S_0$ channel becomes repulsive due to the strong core repulsion in the high density regime~\cite{1966ApJ...145..834W}. The $^3P_2$ interaction stemming from a strong spin-orbit force between nucleons remains attractive, leading to the formation of spin-triplet $p$-wave Cooper pairs preserving  total angular momentum $J=2$. Neutron $^3P_2$ superfluids have been predicted to occupy the neutron-star interior instead of the conventional $^1S_0$ superfluids near the surface~\cite{Tabakin:1968zz,Hoffberg:1970vqj,Tamagaki1970,Hoffberg:1970vqj,takatsukaPTP71,takatsukaPTP72,fujitaPTP72,Richardson:1972xn,Amundsen:1984qc,Takatsuka:1992ga,sauls_nato,Baldo:1992kzz,Elgaroy:1996hp,Khodel:1998hn,Baldo:1998ca,Khodel:2000qw,Zverev:2003ak,Maurizio:2014qsa,Bogner:2009bt,Srinivas:2016kir}. In contrast to $^1S_0$ superfluids, the $^3P_2$ superfluids are tolerant against a strong magnetic field to the spin-triplet pairings. Thus the $^3P_2$ superfluids will be more important in magnetars. It has been discussed that the enhancement of neutrino emission from the onset of the breaking and formation of $^3P_2$ Cooper pairs may explain the observed rapid cooling of the neutron star in Cassiopeia A~\cite{Heinke2010,Shternin2011,Page:2010aw} (see also Refs.~\cite{Blaschke:2011gc,Blaschke:2013vma,Grigorian:2016leu}).

Let us consider neutrons interacting via a strong spin-orbit force, 
which are invariant under a gauge transformation [${\rm U}(1) _{\varphi}$] and spin-momentum rotation [${\rm SO}(3)_{J}$].\footnote{In this paper, we consider the presence of a uniform magnetic field along the $z$-axis, which explicitly breaks ${\rm SO}(3)_J$ down to ${\rm U}(1)_{J_z}$.} The symmetry group, $G= {\rm U}(1) _{\varphi}\times {\rm SO}(3) _{J}$, has three continuous subgroups, ${\rm U}(1)_{J_z}$, ${\rm U}(1)_{2J_z-\varphi}$, and ${\rm U}(1)_{J_z-\varphi}$. The first case is the uniaxial nematic (UN) phase. The UN phase is thermodynamically stable at zero magnetic fields, while a magnetic field induces successive phase transitions from the UN phase to BN phases with the dihedral-two ($D_2$) or dihedral-four ($D_4$) symmetries~\cite{Mizushima:2016fbn,Masuda:2015jka,Masuda:2016vak}. 
These nematic phases appear also in spin-2 Bose-Einstein condensates \cite{kawaguchi}. 
However, one of the most salient features coming from the fact that $^3P_2$ superfluid is composed of fermion Cooper pairs 
is that
all the nematic phases are prototypes of topological superfluidity accompanied by topologically protected Bogoliubov quasiparticles~\cite{Mizushima:2016fbn,Mizushima:2017pma}. 
In addition to these fermionic excitations, there are also
rich massless and massive bosonic excitations~\cite{Bedaque:2003wj,Leinson:2011wf,Leinson:2012pn,Leinson:2013si,Bedaque:2012bs,bedaquePRC14,Bedaque:2013fja,Bedaque:2014zta,Leinson:2009nu,Leinson:2010yf,Leinson:2010pk,Leinson:2010ru,Leinson:2011jr}, which might provide sources of cooling and transport phenomena in neutron stars. 
The ordered states characterized by the other two subgroups, ${\rm U}(1)_{2J_z-\varphi}$ and ${\rm U}(1)_{J_z-\varphi}$, are referred to as the ferromagnetic (FM) phases. These are the condensation of $^3P_2$ Cooper pairs with fully polarized spins. The interaction of neutron spins with strong field in magnetars gives rise to the Zeeman splitting of the neutron Fermi surface, amounting to the critical field of the Pauli depairing of $^1S_0$ Cooper pairs ($\sim k_{\rm B}T_{\rm c}$) or more, where the superfluid critical temperature is about $T_{\rm c}\sim 10^{9}~{\rm K}$. The spin-polarized states are compatible with a strong Zeeman splitting, and may become competitive to nematic states at strong magnetic fields. Nevertheless, no previous works have predicted the thermodynamic stability of spin-polarized superfluid states.  

In this paper, we report the thermodynamic stability of spin-polarized $^3P_2$ superfluid phases, such as the magnetized BN and FM phases, at strong magnetic fields relevant to magnetars. In general, the superfluid phase transition drastically changes a low-energy dispersion of neutrons near the Fermi surface. The Cooper pairs are confined to a low-energy band near the Fermi surface and a shell in momentum space, $|{\bm k}-{\bm k}_{\rm F}|\ll k_{\rm F}$, where ${\bm k}_{\rm F}$ is the Fermi momentum of normal neutrons. In previous works~\cite{Mizushima:2016fbn,Mizushima:2019spl}, therefore, we employed the quasiclassical approximation. 
We used the same approximation 
 in our studies on $^3P_2$ superfluids based on the Ginzburg-Landau (GL) 
 theory 
 \cite{Masuda:2015jka,Masuda:2016vak,
 Yasui:2018tcr,Yasui:2019unp}.
The approximation incorporates quasiparticles confined to a low-energy shell near the Fermi surface, and quasiparticles away from the Fermi level are renormalized to the critical temperature and Fermi liquid parameters. The quasiclassical approximation postulates an infinitely large Fermi surface and cannot properly incorporate the Zeeman splitting of neutron Fermi surface. In this paper, we perform the microscopic calculation of the Luttinger-Ward thermodynamic potential without the quasiclassical approximation. The finite-size correction of the neutron Fermi surface favors the FM state at high temperatures and strong magnetic fields, while the $D_4$-BN phase still survives in low temperatures. The Zeeman splitting does not exclude the $D_4$-BN phase under strong magnetic fields, but induces spin polarization of Cooper pairs, which is referred to as the magnetized BN phase (This is also called a broken axisymmetric phase in spin-2 BEC~\cite{kawaguchi}). We also discuss the impact of such spin polarization on topological excitations, such as Abelian and non-Abelian fractional vortices. 
Furthermore, we extend the GL theory to incorporate the finite-size effect of the Fermi surface. The augmented GL theory properly captures the thermodynamics of the spin-polarized $^3P_2$ superfluid phases.

The organization of this paper is as follows. In Sec.~\ref{sec:theory}, we present the Luttinger-Ward functional for neutron $^3P_2$ superfluids and symmetry classification of $^3P_2$ superfluid phases including UN/BN phases and FM phases. Here we emphasize that the particle-hole asymmetry is indispensable for the thermodynamic stability of the spin-polarized phase. Based on the theory, in Sec.~\ref{sec:phase}, we show that the phase diagram of $^3P_2$ superfluids at strong magnetic fields. In Sec.~\ref{sec:gl}, we present the augmented GL theory which properly takes account of the leading order contributions of the particle-hole asymmetry. Sec.~\ref{sec:summary} is devoted to a summary and discussion.

\section{$^3P_2$ superfluid theory}
\label{sec:theory}

\subsection{Luttinger-Ward theormodynamic functional}
\label{sec:scf}

Here we introduce the theoretical formulation for neutron $^3P_2$ superfluids on the basis of the Luttinger-Ward theormodynamic functional. Let us start to define the Nambu-Gor'kov Green's function in terms of a grand ensemble average of the fermion-field operators in the Nambu (particle-hole) space, ${\bm \Psi}\equiv(\psi_{\uparrow},\psi _{\downarrow}, \bar{\psi}_{\uparrow},\bar{\psi}_{\downarrow})^{\rm tr}$, as 
\beq
\hat{G}(x_1,x_2) = -\langle {\rm T}_{\tau}{\bm \Psi} (x_1)\bar{\bm \Psi}(x_2)\rangle
= \begin{pmatrix}
G(x_1,x_2) & F(x_1,x_2) \\ \bar{F}(x_1,x_2) & \bar{G}(x_1,x_2)
\end{pmatrix}, 
\eeq
where $\psi _a(\tau)=e^{\mathcal{H}\tau}\psi _ae^{-\mathcal{H}\tau}$ and $\bar{\psi}_a(\tau)=e^{\mathcal{H}\tau}\psi^{\dag} _ae^{-\mathcal{H}\tau}$ ($a=\uparrow,\downarrow$ for spins) are the field operators for fermions with spin $\alpha=\uparrow,\downarrow$ and momentum ${\bm k}$ in the Matsubara representation. In this paper, we set $\hbar = k_{\rm B}=1$ and set the volume of the system to $V\rightarrow 1$. We have introduced abbreviation, $x_i\equiv ({\bm r}_i,\tau_i)$, representing the three dimensional space position ${\bm r}_{i}$ and the imaginary time $\tau_{i}$ for the neutron ($i=1,2$). $a^{\rm tr}$ denotes the transpose of a matrix $a$. 
The elements of $\hat{G}$ obey the symmetry relations in the Nambu space, $G_{\alpha\beta}(x_1,x_2) = - G_{\beta\alpha}(x_2,x_1)$ and $\bar{F}_{\alpha\beta}(x_1,x_2) = F^{\ast}_{\beta\alpha}(x_2,x_1)$.
In this paper, we consider translationally invariant neutron matter and transform the space-time position $x$ to the momentum ${\bm k}$ and Matsubara frequency at temperature $T$, $\varepsilon_n = (2n+1)\pi T$ ($n\in\mathbb{Z}$): $(x_1,x_2) \rightarrow ({\bm k},\varepsilon_n)$. The self-consistent formalism is derived from the Luttinger-Ward thermodynamic functional which is given in terms of the full Nambu-Gor'kov Green's function $G$ and the self-energy $\Sigma$ as
\begin{align}
\Omega [\hat{G}] =& -\frac{1}{2}T\sum _{n} \int \frac{d^3k}{(2\pi)^3}{\rm Tr}_4\bigg\{
\hat{\Sigma} \hat{G}
+ \ln \left(
-\hat{G}^{-1}_0 + \hat{\Sigma}
\right)
\bigg\} \nn \\
& +\frac{1}{2} \int \frac{d^3k}{(2\pi)^3} {\rm tr}_2\xi_{\bm k}
+ \Phi[\hat{G}],
\label{eq:LW}
\end{align}
where the trace, ${\rm Tr}_4$, (${\rm tr}_2$) is taken over the spin and Nambu spaces (spin space). The inverse propagator for free fermions is given by \beq
\hat{G}^{-1}_0({\bm k},i\varepsilon_n) = i\varepsilon_n -{\xi} _{\bm k},
\eeq 
where ${\xi} _{\bm k}$ is the $2\times 2$ spin matrix of the single-particle Hamiltonian for neutrons under a magnetic field ${\bm B}$ 
\beq
\xi_{\bm k} = \frac{{\bm k}^{2}}{2m}-\mu-\frac{1}{2}\gamma_{\rm n} {\bm \sigma }\cdot {\bm B},
\label{eq:ek}
\eeq
where $m$ and $\mu$ are the mass and the chemical potential. Here $\gamma_{\rm n}=1.2 \times 10^{-13}$ MeV/T is the gyromagnetic ratio for a neutron.
We use ${\bm \sigma}=(\sigma_{1},\sigma_{2},\sigma_{3})$ and ${\bm \tau}=(\tau_{1},\tau_{2},\tau_{3})$ to denote the matrices in the spin and Nambu spaces, respectively. The last term in Eq.~\eqref{eq:ek} is the interaction of neutron spins with magnetic field. The Green's function and the self-energy are related to the functional $\Phi[\hat{G}]$ by the stationary conditions with respect to the Green's function, $\delta \Omega/\delta \hat{G}^{\rm tr}=0$, and the self-energy, $\delta \Omega/\delta \hat{\Sigma}^{\rm tr}=0$. The former is recast into the definition of the self-energy in terms of the functional derivative
\beq
\hat{\Sigma} [\hat{G},\hat{\Sigma}] = 2 \frac{\delta \Phi[\hat{G}]}{\delta \hat{G}^{\rm tr}}
\equiv \begin{pmatrix}
\Sigma & \Delta \\ \bar{\Delta} & \bar{\Sigma}
\end{pmatrix}.
\label{eq:sigmaLW}
\eeq
The Dyson's equation for the full Green's function is obtained from the latter stationary condition as 
\beq
\hat{G}^{-1}=\hat{G}^{-1}_0 - \hat{\Sigma}[\hat{G}].
\label{eq:dyson}
\eeq



The off-diagonal submatrix of the self-energy, $\Delta_{\alpha\beta}({\bm k})$, in Eq.~\eqref{eq:sigmaLW} is the mean field associated with effective interaction in the Cooper channel of neutrons, 
\begin{gather}
\Delta_{\alpha\beta}({\bm k}) = T\sum_n\int \frac{d^3k^{\prime}}{(2\pi)^3}V^{\rm pair}_{\alpha\beta;\gamma\delta}({\bm k},{\bm k}^{\prime})
F_{\gamma\delta}({\bm k}^{\prime},i\varepsilon_n),
\label{eq:gap}
\end{gather}
where we assume a condensate of neutron pairs with zero center-of-mass energy and momentum. The diagonal self-energy,
\beq
\Sigma_{\gamma\alpha}({\bm k})=T\sum_n\int \frac{d^3k^{\prime}}{(2\pi)^3}V^{\rm FL}_{\alpha\beta;\gamma\delta}({\bm k},{\bm k}^{\prime})
G_{\beta\delta}({\bm k}^{\prime},i\varepsilon_n),
\eeq
describes the Fermi liquid corrections to the quasiparticle excitation energies. The $\Phi$-functional in Eq.~\eqref{eq:LW} are constructed from these self-energies as 
\begin{align}
\Phi =& \frac{1}{4}T\sum_n\int \frac{d^3k}{(2\pi)^3} 
T\sum_{n^{\prime}}\int \frac{d^3k^{\prime}}{(2\pi)^3} \nn \\
& \times \left[ G_{\gamma\alpha}({\bm k},i\varepsilon_n)V^{\rm FL}_{\alpha\beta;\gamma\delta}({\bm k},{\bm k}^{\prime})
G_{\beta\delta}({\bm k}^{\prime},i\varepsilon_{n^{\prime}})  \right.\nn \\
&\left. +2\bar{F}_{\alpha\beta}({\bm k},i\varepsilon_n) 
V^{\rm pair}_{\alpha\beta;\gamma\delta}({\bm k},{\bm k}^{\prime}){F}_{\gamma\delta}({\bm k}^{\prime},i\varepsilon_{n^{\prime}})\right].
\end{align}

The interaction in the Cooper channel is decomposed into the spin-singlet (e: even parity) and spin-triplet (o: odd parity) functions:
$V^{\rm pair}_{\alpha\beta;\gamma\delta}(\hat{\bm k},\hat{\bm k}^{\prime}) = 
V^{({\rm e})}_{\alpha\beta;\gamma\delta}(\hat{\bm k},\hat{\bm k}^{\prime})
+V^{({\rm o})}_{\alpha\beta;\gamma\delta}(\hat{\bm k},\hat{\bm k}^{\prime})
$, where we have introduced $\hat{\bm k}\equiv {\bm k}/k_{\rm F}$. 
At low density of neutrons, the $^1S_0$ channel is attractive and dominant in Cooper pair channels. At the high density regime $\rho \gtrsim 10^{14}{\rm g}/{\rm cm}^{3}$ ($\rho$ is the particle density of neutrons), the $^1S_0$ channel remains repulsive due to the strong core repulsion, while the $^3P_2$ interaction stemming from a strong spin-orbit force between two nucleons remains
 attractive. 
As the interaction in the $^{3}P_{0}$ and $^{3}P_{1}$ channels are repulsive at high density~\cite{Hoffberg:1970vqj,Tamagaki1970,Dean:2002zx}, they are irrelevant to the formation of the superfluidity. Hence we consider the attractive $^3P_2$ interaction as a dominant pairing interaction of neutrons in the inner cores of neutron stars, 
\beq
V^{\rm pair}_{\alpha\beta;\gamma\delta}(\hat{\bm k},\hat{\bm k}^{\prime})
= -v \sum_{\mu,\nu=1}^{3} T_{\mu\nu,\alpha\beta}(\hat{\bm k})T^{\ast}_{\mu\nu,\delta\gamma}(\hat{\bm k}^{\prime}),
\eeq
The traceless and symmetric tensor, $T_{\mu\nu,\alpha\beta}(\hat{\bm k})$, is given by
\begin{align}
T_{\mu\nu,\alpha\beta}(\hat{\bm k})
= 
\frac{1}{2}\left( 
\Lambda^{\mu}_{\alpha\beta} \hat{k}_{\nu} + \Lambda^{\nu}_{\alpha\beta} \hat{k}_{\mu}
\right) - \frac{1}{3}\delta _{\mu\nu}{\bm \Lambda}_{\alpha\beta}\!\cdot\! \hat{\bm k},
\end{align}
with $\Lambda^{\mu}_{\alpha\beta} = i(\sigma_{\mu}\sigma_y)_{\alpha\beta}$
which obeys $T_{\mu\nu,\alpha\beta}(\hat{\bm k}) = T_{\nu\mu,\alpha\beta}(\hat{\bm k}) = T_{\mu\nu,\beta\alpha}(\hat{\bm k}) $ and ${\rm tr}(T(\hat{\bm k})) \equiv\sum _{\mu}T_{\mu\mu,\alpha\beta}(\hat{\bm k})=0$~\cite{Richardson:1972xn}. 
The general form of the spin-triplet $p$-wave superfluid order is expressed in terms of a symmetric traceless tensor, $\mathcal{A}_{\mu i}\!\in\!\mathbb{C}$ as 
\beq
{\Delta}_{\alpha\beta}({\bm k}) 
= \sum _{\mu\nu} \mathcal{A}_{\mu\nu}T_{\mu\nu,\alpha\beta}(\hat{\bm k}).
\label{eq:A}
\eeq

Another interaction potential, ($V^{\rm FL}_{\alpha\beta;\gamma\delta}$), is expanded in terms of the Legendre polynomials $P_{\ell}$ and parameterized with the Landau's spin-independent and spin-dependent Fermi-liquid parameters $F^{({\rm s})}_{\ell}$ and $F^{({\rm a})}_{\ell}$ ($\ell=0,1,\dots$) as 
\begin{align}
V^{\rm FL}_{\alpha\beta;\gamma\delta}(\hat{\bm k},\hat{\bm k}^{\prime}) = 
N^{-1}_{\rm F}&\sum _{\ell}\left[ F^{({\rm s})}_{\ell}\delta _{\alpha\gamma}\delta_{\beta\delta}
+F^{({\rm a})}_{\ell}{\bm \sigma}_{\alpha\gamma}\!\cdot\!{\bm \sigma}_{\beta\delta}
\right]\nn \\ 
&\times P_{\ell}(\hat{\bm k}\!\cdot\!\hat{\bm k}^{\prime}),
\end{align}
where $N_{\rm F}$ is the neutron density of states at the Fermi energy $\varepsilon_{\rm F}$.
Among the dimensionless parameters, $F^{({\rm a})}_0$ gives rise to the spin-exchange interaction with the magnetization density of neutrons, and significantly affects the equilibrium properties of $^3P_2$ superfluids under a magnetic field. In this work, therefore, we only take account of the lowest order terms ($\ell=0$). The $2\times 2$ self-energy matrix in the spin space is then given by 
\beq
{\Sigma}_{\alpha\beta} = \frac{1}{2N_{\rm F}} \left[ F^{({\rm s})}_0 n \delta _{\alpha\beta} + F^{({\rm a})}_0{\bm S}\!\cdot\! 
{\bm \sigma}_{\alpha\beta}\right],
\eeq
and $\bar{\Sigma}_{\alpha\beta}=-\Sigma_{\beta\alpha}$.
The particle density $n$ and the spin density ${\bm S}$ are obtained as 
\begin{gather}
n = T\sum _n \int \frac{d^3k}{(2\pi)^3} {\rm tr}_2[{G}({\bm k},i\varepsilon_n)], \\
{ S}_{\mu} = T\sum _n \int \frac{d^3k}{(2\pi)^3}{\rm tr}_2\left[
{ \sigma}_{\mu} {G}({\bm k},i\varepsilon_n)\right],
\end{gather}
respectively.
Although the thermodynamic properties are insensitive to $F^{({\rm s,a})}_{\ell\ge 1}$, 
the fermionic vacuum polarization steaming from the Fermi liquid corrections with $\ell \ge 1$ may affect the spectrum of bosonic excitations, i.e., collective excitations of $^3P_2$ Cooper pairs~\cite{saulsPRB17}. 

The full Green's function, $\hat{G}$, is determined from the self-consistency condition in Eq.~\eqref{eq:sigmaLW} as 
\beq
\hat{G}^{-1}({\bm k},i\varepsilon_n) = i\varepsilon_n - \hat{\mathcal{H}}_{\rm BdG}({\bm k}), 
\eeq
where the Bogoliubov-de Gennes Hamiltonian is given by 
\beq
\hat{\mathcal{H}}_{\rm BdG}({\bm k})\equiv \begin{pmatrix}
{\xi}_{\rm eff}({\bm k}) & {\Delta}({\bm k}) \\ -{\Delta}^{\ast}(-{\bm k}) & -{\xi}^{\rm tr}_{\rm eff}(-{\bm k}) 
\end{pmatrix}.
\eeq
The self-energies associated with the particle density and spin density are renormalized to the chemical potential and the magnetic field as 
\beq
{\mu}_{\rm eff} = \mu - \frac{1}{2N_{\rm F}}F^{({\rm s})}_0 n, \quad
{\bm B}_{\rm eff} \equiv {\bm B} - \frac{F^{({\rm a})}_0{\bm M}}{\chi^{0}_{\rm N}},
\label{eq:mu_B}
\eeq
where $M_{\mu}\equiv \frac{1}{2}\gamma_{\rm n}S_{\mu}$ is the magnetization density of neutron $^3P_2$ superfluid and $\chi^0_{\rm N}$ is the Pauli spin susceptibility of a normal neutron gas with effective mass $m^{\ast}$ and the density of state $N_{\rm F}$. 
We define ${\xi}_{\rm eff}({\bm k})={\bm k}^{2}/(2m)-\mu_{\rm eff}-({1}/{2})\gamma_{\rm n} {\bm \sigma }\cdot {\bm B}_{\rm eff}$.
Let $\hat{\mathcal{U}}({\bm k})$ be a $4\times 4$ unitary matrix that diagonalizes the Bogoliubov-de Gennes Hamiltonian as 
\beq
\hat{\mathcal{U}}^{\dag}({\bm k})\hat{\mathcal{H}}_{\rm BdG}({\bm k})\hat{\mathcal{U}}({\bm k})=\begin{pmatrix}
{\bf E}({\bm k}) & \\ & -{\bf E}(-{\bm k})
\end{pmatrix},
\label{eq:bdg}
\eeq
where ${\bf E}({\bm k})={\rm diag}(E_{1}({\bm k}),E_{2}({\bm k}))$ is the energy eigenvalues of the Hamiltonian. The thermodynamic stability of $^3P_2$ superfluids is determined by minimizing the thermodynamic potential $\Omega(T,\mu,B)$. The potential is equivalent to the Luttinger-Ward functional $\Omega[\hat{G},\hat{\Sigma}]$ when $\hat{G}$ and $\hat{\Sigma}$ satisfy the stationary conditions in Eqs.~\eqref{eq:sigmaLW} and \eqref{eq:dyson}. On the basis of the quasiparticle energy, $E_{\alpha}({\bm k})$, the thermodynamic potential is recast into 
\begin{align}
\Omega
=& \frac{1}{2}\sum_{\alpha}\int \frac{d^3k}{(2\pi)^3}\left[\xi_{\alpha}({\bm k})-E_{\alpha}({\bm k})-2T\ln(1+e^{-E_{\alpha}({\bm k})/T})\right] \nn \\
&+\frac{1}{2{v}}{\rm tr}_3[\mathcal{A}^{\dag}\mathcal{A}]-\frac{1}{4N_{\rm F}}F^{({\rm s})}_0n^2 + \frac{1}{4N_{\rm F}}F^{({\rm a})}_0S^2_z,
\label{eq:omegaMF}
\end{align}
where we utilize the relations, ${\rm tr}\ln A = \ln \det A$ and 
$\sum_n \ln[(\varepsilon^2_n+\varepsilon^2)/T^2]
= {\varepsilon}/T + 2\ln ( 1 + e^{-\varepsilon/T})+a$ ($a$ is a $T$-independent constant). 
The gap equation for the $^3P_2$ order parameter are derived from the stationary condition of the thermodynamic potential.
The total particle density $n$ and magnetization density $M_{\mu}=\frac{1}{2}\gamma_{\rm n}S_{\mu}$ are also obtained from the thermodynamic relations 
\begin{gather}
n=-\frac{\partial \Omega}{\partial \mu}, 
\quad
M_{\mu}=-\frac{\partial \Omega}{\partial B_{\mu}},
\label{eq:nm}
\end{gather}
respectively. The gap equations for $\mathcal{A}_{\mu i}$ are derived from the stationary condition of the thermodynamic potential, 
\beq
\frac{\delta\Omega}{\delta \mathcal{A}^{\ast}_{\mu i}}=0. \label{eq:gapequation}
\eeq
The equilibrium state is determined by solving the self-consistent equations 
\eqref{eq:mu_B}, \eqref{eq:nm}, and the gap equation \eqref{eq:gapequation} for $\mathcal{A}_{\mu i}$.

\subsection{Symmetry classification of $^3P_2$ order parameters}
\label{sec:symmetry}

In the absence of external fields, the normal neutron system is invariant under a gauge transformation [${\rm U}(1)_{\varphi}$] and the three-dimensional spin-orbit rotation [${\rm SO}(3)_J$]
\beq
G={\rm U}(1)_{\varphi} \times {\rm SO}(3)_J.
\eeq
The order parameter for $^3P_2$ superfluids is the $3\times 3$ traceless symmetric tensor, $\mathcal{A}_{\mu i}$, defined in Eq.~\eqref{eq:A}. The continuous symmetry, $G$, acts on the tensor as 
\beq
\mathcal{A} \rightarrow e^{i\varphi}g\mathcal{A}g^{\rm tr} , \quad e^{i\varphi}\in {\rm U}(1),~g\in{\rm SO}(3)
\eeq
in the $3\times 3$ matrix notation. The ordered state spontaneously breaks the symmetry of the Hamiltonian, $G$, to $H\subset G$. The group $H$ describes the remaining symmetry of the ordered state and the tensor is invariant under a transformation by the group element $h\in H$, $h\mathcal{A}=\mathcal{A}$. The element $h$ is written as $h=e^{i\Phi + i\alpha_{\mu} {\sf J}_{\mu}}$, where $\Phi$ and ${\sf J}_{\mu}$ are the generators for a gauge transformation and simultaneous spin-orbit rotation, respectively, and ${\bm \alpha}$ is an infinitesimal parameter. 

There are three continuous subgroups of $G$, $H_{\rm FM}={\rm U}(1)_{\varphi-2J_z}$, $H_{\rm FM^{\prime}}={\rm U}(1)_{\varphi-J_z}$, and $H_{\rm UN}={\rm U}(1)_{J_z}$. The elements of these subgroups are given by $h=e^{i(\alpha-2{\sf J}_z\alpha)}$, $h=e^{i(\alpha-{\sf J}_z\alpha)}$, and $h=e^{i{\sf J}_z\alpha}$, respectively. In the first two subgroups, the gauge transformation $\mathcal{A}\rightarrow \mathcal{A}e^{i\alpha}$ is compensated by the spin-orbit rotation about the $z$ axis by $-\alpha$ or $-2\alpha$, which manifestly exhibits the equivalence between phase change and spin-orbit rotation. The order parameters of two ferromagnetic (${\rm FM}$ and ${\rm FM}^{\prime}$) phases are given by 
\begin{align}
\mathcal{A}^{\rm FM}_{\mu i} = \Delta\begin{pmatrix}
1 & \pm i & 0 \\ \pm i & -1 & 0 \\ 0 & 0 & 0
\end{pmatrix}_{\mu i}, \quad
\mathcal{A}^{{\rm FM}^{\prime}}_{\mu i} = \Delta\begin{pmatrix}
0 & 0 & 1 \\ 0 & 0 & \pm i \\ 1 & \pm i & 0
\end{pmatrix}_{\mu i},
\label{eq:FM}
\end{align}
where we omit the ${\rm U}(1)$ phase degrees of freedom. The former (latter) state corresponds to a condensation of neutrons into Cooper pairs with $J_z=\pm 2$ ($J_z=\pm 1$), whose spins are fully (partially) polarized. The ordered state with $H_{\rm UN}={\rm U}(1)_{J_z}$ is the UN nematic state.

In addition to the ordered states preserving continuous symmetry, there exist competitive orders with discrete symmetry. When the ordered state preserves the time-reversal symmetry, $\mathcal{A}_{\mu i}$ must be a real symmetric traceless tensor. Hence, the order parameter with time-reversal symmetry is given by a diagonal form
\begin{align}
\mathcal{A}_{\mu i} &= \Delta\begin{pmatrix}
1 & 0 & 0 \\ 0 & r & 0 \\ 0 & 0 & -1-r
\end{pmatrix}_{\mu i}.
\end{align}
The eigenvalue $r\in[-1,-1/2]$ is the order parameter that represents the biaxiality of the nematic order parameter. The most symmetric state is the UN state at $r=-1/2$, which maintains the ${\rm U}(1)$ symmetry about the $\hat{x}$ axis in the spin-orbit space.
When $r$ deviates from the UN point, the ordered state breaks the ${\rm U}(1)$ symmetry down to the discrete symmetry. The BN state at $r=-1$ remains invariant under dihedral-four $D_4$ symmetry, which has the $C_4$ and $C_2$ axes. The intermediate $r$ holds dihedral-two $D_2$ symmetry with the three $C_2$ axes. 

Below, we focus on the high temperature and high magnetic field regime of $^3P_2$ neutron superfluids. The $^3P_2$ order parameter under  a magnetic field is expressed in terms of three real parameters $(\Delta,\kappa,r)$ as 
\begin{align}
\mathcal{A}_{\mu i} &= \Delta\begin{pmatrix}
1 & i\kappa & 0 \\ i\kappa & r & 0 \\ 0 & 0 & -1-r
\end{pmatrix}_{\mu i}.
\label{eq:OP}
\end{align}
Without loss of generality, we assume that the magnetic field is applied along the $z$-axis.
The order parameter in Eq.~\eqref{eq:OP} has three real variational parameters, $\Delta$, $r\in[-1,-1/2]$, and $\kappa\in[-1,1]$, which are determined by solving the self-consistent equations in Sec.~\ref{sec:scf}. The amplitude of the order parameter is represented by $\Delta$, while $r\in[-1,-1/2]$ and $\kappa\in[-1,1]$ quantify the biaxiality and spin-polarization of the nematic order parameter, respectively. The order parameter in Eq.~\eqref{eq:OP} contains the FM states in $(r=-1,\kappa=-1)$ and nonmagnetic UN/BN states in $\kappa=0$.

{\renewcommand\arraystretch{1.4}}
\begin{table*}[t!]
\caption{The remaining symmetries ($H$), order parameter manifolds ($R=G/H$) and topological vortices ($\pi_1(R)$) in possible phases of $^3P_2$ superfluids, the uniaxial/biaxial nematic phases, the cyclic phase, the magnetized BN phase, and the ferromagnetic phase. The magnetized $D_4$-BN phase with $r=-1$ and $\kappa \neq 0$ is also called a broken axisymmetric phase in spin-2 BEC~\cite{kawaguchi}. $\#_{\rm NG}$ and $\#_{\rm qNG}$ count the number of Nambu-Goldstone (NG) modes and quasi-Nambu-Goldstone (qNG) modes, respectively. $A^{\ast}$ indicates the universal covering group of $A$, and $\mathbb{Q}=D^{\ast}_2$ is a quaternion group (see Appendix A of Ref.~\cite{Kobayashi:2011xb} for the definition of the product $\times _h$).}
\begin{ruledtabular}
\begin{tabular}{ccccccccc}
& Phase & O.P. [see Eq.~\eqref{eq:OP}] & $H$ & $R=G/H$ & $\pi _1 (R)$ & $\#_{\rm NG}$ & $\#_{\rm qNG}$~\cite{uchino} & \\
\hline
& Uniaxial nematic & $r=-1/2$, $\kappa=0$ & {$D_{\infty} \simeq {\rm O}(2)$} & ${\rm U}(1)\times \mathbb{R}P^2$ & $\mathbb{Z}\oplus\mathbb{Z}_2$~\cite{song,Masuda:2015jka} & 3 & 2 &\\
\hline
& Biaxial nematic & \begin{tabular}{c} $r\in (-1,-1/2)$, $\kappa=0$ \\ $r=-1$, $\kappa=0$ \end{tabular} & \begin{tabular}{c} $D_2$ \\ $D_4$ \end{tabular} & 
\begin{tabular}{c} ${\rm U}(1)\times {\rm SO}(3)/D_2$ \\ $[{\rm U}(1)\times {\rm SO}(3)]/D_4$ \end{tabular} & 
\begin{tabular}{c} $\mathbb{Z}\oplus\mathbb{Q}$~\cite{song,Masuda:2015jka} \\ $\mathbb{Z}\times _h D_4^{\ast}$~\cite{Masuda:2015jka,Masuda:2016vak,Kobayashi:2011xb} \end{tabular} & 
\begin{tabular}{c} $4$ \\ $4$ \end{tabular} &  \begin{tabular}{c} $1$ \\ $1$ \end{tabular} &\\
\hline
& Cyclic & $r=e^{i2\pi /3}$, $\kappa=0$ & $T$ & $[{\rm U}(1)\times {\rm SO}(3)]/T$ & $\mathbb{Z}\times _h T^{\ast}$~\cite{semenoff,makela,Kobayashi:2008pk,Kobayashi:2011xb} & 3 & --- &\\ 
\hline
& \begin{tabular}{c} Magnetized \\ biaxial nematic
\end{tabular} &\begin{tabular}{c} $r\in (-1,-1/2)$, $\kappa\in(0,1)$ \\ $r=-1$, $\kappa\in(0,1)$ \end{tabular} & \begin{tabular}{c} 0 \\ $C_4$ \end{tabular} & 
\begin{tabular}{c} ${\rm SO}(3)\times {\rm U}(1)$ \\ $[{\rm U}(1)\times{\rm SO}(3)]/\mathbb{Z}_4$ \end{tabular} & 
\begin{tabular}{c} $\mathbb{Z}_2\oplus\mathbb{Z}$ \\ $\mathbb{Z}\times _h C_4^{\ast}$ \end{tabular} & \begin{tabular}{c} $4$ \\ {$4$} \end{tabular} & 
\begin{tabular}{c} --- \\ --- \end{tabular} & \\
\hline
& Ferromagnetic & \begin{tabular}{c} $r=-1$, $\kappa=1$ \\ Eq.~\eqref{eq:FM} \end{tabular} & \begin{tabular}{c} ${\rm U}(1)_{J_z+2\Phi}$ \\
{${\rm U}(1)_{J_z+\Phi}$} \end{tabular} & 
\begin{tabular}{c} ${\rm SO}(3)_{J_z-2\Phi}/\mathbb{Z}_2$ \\ {${\rm SO}(3)_{J_z-\Phi}/\mathbb{Z}_2$} \end{tabular}& 
\begin{tabular}{c} $\mathbb{Z}_4$~\cite{makela,vollhardt2013superfluid} \\ {$\mathbb{Z}_4$~\cite{makela,vollhardt2013superfluid}} \end{tabular} & 
\begin{tabular}{c} 3 \\ 3 \end{tabular} & \begin{tabular}{c} --- \\ --- \end{tabular} &\\ 
\end{tabular}
\end{ruledtabular}
\label{table}
\end{table*}

For $r=-1$ and $\kappa\in(0,1)$ in Eq.~\eqref{eq:OP}, the order parameter is invariant under the cyclic group $C_4\subset G$, whose elements $(e^{i\alpha},g)$ are 
\beq
C_4 = \{ (1,{\bm 1}_3), (-1,R_3),  (1, I_3), (-1,I_3R_3) \},
\eeq
where $I_3$ and $R_3$ represent $\pi$ and $\pi/2$ rotations around the third ($z$) axis, respectively, 
\beq
I_3 = \begin{pmatrix}
-1 & 0 & 0 \\ 0 & -1 & 0 \\ 0 & 0 & 1 
\end{pmatrix}, \quad 
R_3 = \begin{pmatrix}
0 & -1 & 0 \\ 1 & 0 & 0 \\ 0 & 0 & 1 
\end{pmatrix},
\eeq
and ${\bm 1}_3$ is the $3\times 3$ unit matrix. The mixing of nonzero $\kappa$ component in the $D_4$-BN state lowers the symmetry to the cyclic group $C_4$. 

\begin{figure}[b!]
\includegraphics[width=80mm]{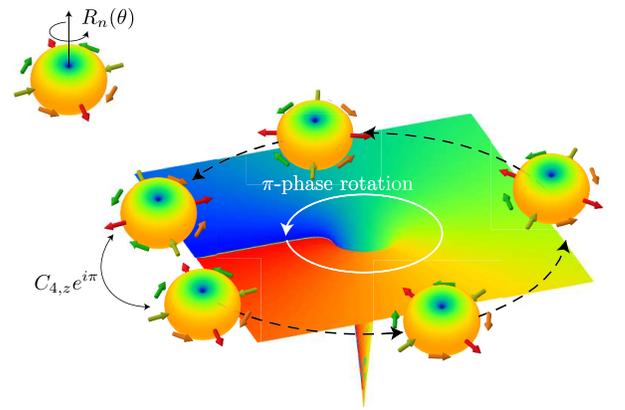}
\caption{Schematics of the texture of the gap structure and ${\rm U}(1)$ phase in a half-quantized vortex of the unitary ($D_4$) BN state. The gap structure shows the energy gap $E_g$ at ${\bm k}={\bm k}_{\rm F}$ and the thick arrows denote the spin degrees of freedom of $^3P_2$ Cooper pairs, ${\bm d}({\bm k})$.}
\label{fig:HQV}
\end{figure}

In Table~\ref{table}, we summarize the remaining symmetries ($H$), 
order parameter spaces ($R=G/H$) and topological (vortex) excitations ($\pi_1(R)$) in possible phases of $^3P_2$ superfluids. In the nonmagnetic BN state with $r=-1$ and $\kappa=0$, the unbroken symmetry is the dihedral-four $D_4$, and the the fundamental group is  $\pi_1(R) = \mathbb{Z}\times _h D_4^{\ast}$, which include the possibility of non-Abelian half-quantized vortices
\cite{Masuda:2015jka,Masuda:2016vak}. Recently, the thermodynamic stability of non-Abelian half-quantized vortices has been discussed by means of the Ginzburg-Landau theory~\cite{Masuda:2016vak} and microscopic quasiclassical theory~\cite{masaki}. In the magnetized BN state with $r=-1$ and $0 < \kappa <1$, however, the $D_4$ symmetry reduces to $C_4$. The fundamental group, $\pi_1(R) = \mathbb{Z}\times _h C_4^{\ast}$, indicates that possible vortex excitations include half-quantized vortices. In Fig.~\ref{fig:HQV}, we illustrate the gap structure and ${\rm U}(1)$ phase in a half-quantized vortex of the nonmagnetic $D_4$-BN state. The $\pi$ phase jump is compensated by the four-fold rotation about the $z$-axis, $C_{4,z}$. In the magnetized BN state, the element $(-1,R_3)$ ensures the topological stability of the half-quantized vortices. The spatial profile of the order parameter along the azimuthal angle $\theta \in [0,2\pi)$ is represented as
\beq
\mathcal{A}_{\mu i}(\theta) = \Delta e^{i {k} \theta}R_n(\theta) \mathcal{A}_{\mu i} 
R^{\rm tr}_n(\theta)
\eeq
where $R_n(\theta)\in {\rm O}(2)$ is the $n$-fold rotation matrix about the $z$-axis,
\beq
R_n(\theta) = \begin{pmatrix}
\cos(n\theta) & -\sin(n\theta) & 0 \\ \sin(n\theta) & \cos(n\theta) & 0 \\ 0 & 0 & 1
\end{pmatrix}.
\eeq 
The integer vortex is ${k} = 1$ and $n=0$, while the half-quantized vortex is ${k} = 1/2$ and $n=\pm 1/4$. 
They obey Abelian exchange statistics. In the case of the nonmagnetic $D_4$-BN ($r=-1,\kappa=0$), each non-Abelian half-quantized vortex hosts a single zero energy mode, which behaves as a non-Abelian anyon~\cite{masaki}. The non-Abelian anyon is protected by topological invariants, such as the mirror Chern number. The nonzero $\kappa$ induces the spin polarization of the BN order and changes the order parameter manifold $R$, but does not affect the topological stability of the non-Abelian anyon.

\section{$^3P_2$ superfluid phase diagram}
\label{sec:phase}

\subsection{Particle-hole asymmetry and Zeeman splitting of transition temperatures}
\label{sec:PHS}

At zero magnetic fields, the excitation energy of Bogoliubov quasiparticles is obtained from Eq.~\eqref{eq:bdg} as 
\beq
E_{\alpha=\pm}({\bm k}) = \sqrt{[\xi^{(0)}_{\rm eff}({\bm k})]^2+|{\bm d}({\bm k})|^2\pm |{\bm d}({\bm k})\times {\bm d}^{\ast}({\bm k})|},
\label{eq:EQP}
\eeq
where $\xi^{(0)}_{\rm eff}({\bm k})={\bm k}^{2}/(2m)-\mu_{\rm eff}$ is the energy of the normal neutrons at zero magnetic fields and the ${\bm d}$-vector is defined as 
\beq
{d}_{\mu}({\bm k})\equiv \frac{1}{2}{\rm tr}_2\left[ -i\sigma_y\sigma_{\mu}{\Delta}({\bm k})\right] = \mathcal{A}_{\mu j}\hat{k}_j,
\eeq 
which represents the spin degrees of freedom of $^3P_2$ Cooper pairs. When the time-reversal symmetry is preserved ($\kappa=0$), the ${\bm d}$-vector is real and the eigenstates are doubly degenerate as a Kramers doublet. A state with nonzero $\kappa$ breaks time-reversal symmetry, and the energy gap of the fermionic excitations at ${\bm k}={\bm k}_{\rm F}$ is determined by 
\beq
E_{\rm g}\equiv \sqrt{{\rm tr}_2[\hat{\Delta}({\bm k}_{\rm F})\hat{\Delta}^{\dag}({\bm k}_{\rm F})]/2}
=\sqrt{|{\bm d}({\bm k}_{\rm F})|^2\pm |{\bm d}({\bm k}_{\rm F})\times {\bm d}^{\ast}({\bm k}_{\rm F})|}.
\eeq 
The ordered state with ${\bm d}({\bm k})\times {\bm d}^{\ast}({\bm k})\neq {\bm 0}$ is referred to as a nonunitary state. In such nonunitary state, the $^3P_2$ Cooper pairs at the point ${\bm k}$ have a spin~\cite{leggett}
\beq
{\bm S}_{\rm pair}({\bm k}) = i {\bm d}({\bm k}) \times {\bm d}^{\ast}({\bm k}).
\eeq
The spin polarization of $^3P_2$ Cooper pairs splits the degeneracy of Bogoliubov excitations at ${\bm k}$, leading to two distinct excitation gaps, $|{\bm d}({\bm k})|^2\pm |{\bm d}({\bm k})\times {\bm d}^{\ast}({\bm k})|$. In particular, when $\kappa\neq 0$, the order parameter in Eq.~\eqref{eq:OP} has a net average spin on the Fermi surface, 
\beq
\langle S^z_{\rm pair} \rangle = 2\kappa (1-r) \Delta^2/3, 
\eeq
where $\langle \cdots \rangle$ is the average over the neutron Fermi surface. The net spin of $^3P_2$ Cooper pairs leads to the polarization of neutron spins ${\bm S}$ at zero magnetic fields. The spin polarization of $^3P_2$ Cooper pairs may stabilize the nonunitary states with $\kappa\neq 0$, such as the magnetized $D_4$-BN and FM states, under a strong magnetic field.

\begin{figure}[t!]
\includegraphics[width=85mm]{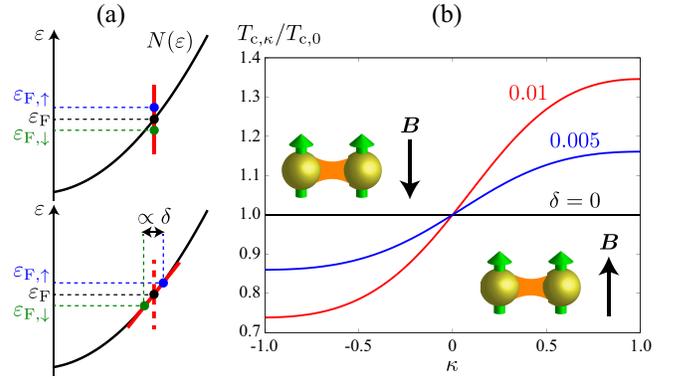}
\caption{(a) Schematics of the particle-hole symmetry emergent in the quasiclassical approximation (top) and the leading-order corrections of the particle-hole asymmetry (bottom). The latter corresponds to the finite size corrections of the Fermi surface, and properly takes account of the Zeeman splitting of the Fermi surface. (b) Critical temperatures as functions of $\kappa$ and $\delta$, where the former (latter) represents the spin polarization of the $^3P_2$ Cooper pairs (the Zeeman splitting of the Fermi surface due to the particle-hole asymmetry). Here we fix $r=-1$.}
\label{fig:splitting}
\end{figure}

In the previous works~\cite{Mizushima:2016fbn,Mizushima:2019spl}, employing the quasiclassical approximation, we found that the UN state is thermodynamically stable at zero magnetic fields, while the phase diagram in nonzero magnetic fields is occupied by the $D_2$ and $D_4$-BN states with $\kappa = 0$. 
The superfluid phase transition drastically changes the properties of normal neutrons, but the change occurs in a low-energy dispersion near the Fermi surface. The Cooper pairs are confined to a low-energy band near the Fermi surface $|\xi_{\rm eff}|\sim \Delta \ll \varepsilon_{\rm F}$ and a shell in momentum space, $|{\bm k}-{\bm k}_{\rm F}|\ll k_{\rm F}$, where $\varepsilon_{\rm F}$ is the Fermi energy of normal neutrons. Hence, the quasiclassical approximation incorporates only quasiparticles confined to a low-energy shell near the Fermi surface as a leading order contribution of $T_{\rm c}/T_{\rm F} \ll 1$ and $\Delta/\varepsilon_{\rm F}\ll 1$.
 The quasiparticles away from the Fermi level are renormalized to the physical parameters, such as the effective mass of neutrons, spin susceptibility, the critical temperature, and so on. 
However, the quasiclassical approximation cannot properly describe the thermodynamic stability of the spin-polarized superfluid states. This is because, as shown in Fig.~\ref{fig:splitting}(a), the approximation assumes the infinitely large Fermi surface and the particle-hole symmetry in the quasiparticle density of states. The self-consistent equations in Sec.~\ref{sec:scf} are thus computed with   
\beq
\int \frac{d^3k}{(2\pi)^3} \approx N_{\rm F} \int d\varepsilon \langle \cdots \rangle ,
\eeq
which cannot incorporate the Zeeman splitting of the neutron Fermi levels and asymmetry in density of states, where $\langle \cdots \rangle$ is the average over the neutron Fermi surface.

To properly discuss the thermodynamic stability of spin-polarized $^3P_2$ superfluids under strong magnetic fields, we compute the momentum integral in self-consistent equations in Sec.~\ref{sec:scf} without employing the quasiclassical approximation. This incorporates the finite-size effect of the Fermi surface and asymmetry of density of states (particle-hole asymmetry), which are regarded as the higher order corrections on $T_{\rm c}/T_{\rm F}$ and $\Delta/\varepsilon_{\rm F}$. It is demonstrated below that the corrections split the critical temperatures of the nonmagnetic UN/BN states and the FM states, and thermodynamically stabilize the spin-polarized $^3P_2$ states under a strong magnetic field.

To see the role of particle-hole asymmetry on the splitting of the critical temperatures, we start with the order parameter in Eq.~\eqref{eq:OP}, which connects the nonmagnetic UN/BN states ($\kappa=0$) and the FM state ($\kappa = \pm 1$). We first note that the critical temperatures in $D_4$-BN and FM states are the same in the quasiclassical approximation with accidental particle-hole symmetry. A strong magnetic field then gives rise to Pauli depairing of $\Delta_{\uparrow\downarrow}=\Delta_{\downarrow\uparrow}$, and favors the ordered state $\Delta_{\uparrow\uparrow}$ and $\Delta_{\downarrow\downarrow}$ with $r = -1$. Then, the order parameter in the spin representation is given as 
\begin{gather}
\Delta_{\uparrow\uparrow}({\bm k}) = -{\Delta}(1+\kappa)(\hat{k}_x+i\hat{k}_y), 
\label{eq:D41} \\
\Delta_{\downarrow\downarrow}({\bm k}) = -{\Delta}(1-\kappa)(\hat{k}_x-i\hat{k}_y), 
\label{eq:D42}
\end{gather}
and $\Delta_{\uparrow\downarrow}({\bm k}) = \Delta_{\downarrow\uparrow}({\bm k})  = 0$. The BdG Hamiltonian is decomposed to the spin sectors
\beq
\hat{\mathcal{H}}_{\rm BdG}({\bm k}) = 
\begin{pmatrix}
{\xi}_{\uparrow}({\bm k}) & {\Delta}_{\uparrow\uparrow}({\bm k}) \\ {\Delta}^{\ast}_{\uparrow\uparrow}({\bm k}) & -{\xi}_{\uparrow}({\bm k}) 
\end{pmatrix} \oplus 
\begin{pmatrix}
{\xi}_{\downarrow}({\bm k}) & {\Delta}_{\downarrow\downarrow}({\bm k}) \\ {\Delta}^{\ast}_{\downarrow\downarrow}({\bm k}) & -{\xi}_{\downarrow}({\bm k}) 
\end{pmatrix},
\eeq
where $\xi _{\uparrow}$ and $\xi_{\downarrow}$ are the single-particle energies of spin-up and down neutrons, respectively. The Bogoliubov quasiparticle excitation energy is obtained as
\beq
E_{\alpha}({\bm k}) = \sqrt{\xi _{\alpha}^2 ({\bm k}) + \Delta^2(1+\alpha\kappa)^2(\hat{k}^2_x+\hat{k}^2_y)}, 
\eeq
where we have taken $\alpha=+$ ($-$) for spin-up (down).

The variational parameters $\kappa$ and $\Delta$ are determined by solving the gap equation. 
{From the saddle point condition of the thermodynamic potential for ${\mathcal A}_{\mu i}$ given by Eq.~\eqref{eq:OP}}, the equations for $\kappa$ and $\Delta$ are obtained as 
\begin{gather}
\frac{1}{{v}} = \frac{1}{2}\left[\frac{(1+ \kappa)^2}{1+\kappa^2}\mathcal{F}_{+} + 
\frac{(1- \kappa)^2}{1+\kappa^2}\mathcal{F}_{-}\right], 
\label{eq:gap1} \\
\frac{\kappa}{{v}} = \frac{1}{2}\left[(1+ \kappa)\mathcal{F}_{+}
- (1-\kappa)\mathcal{F}_{-}
\right],
\label{eq:gap2}
\end{gather}
where
\beq
\mathcal{F}_{\alpha} = \sum _{\bm k}\frac{\hat{k}^2_x+\hat{k}^2_y}{2E_{\alpha}({\bm k})}\tanh\left(
\frac{E_{\alpha}({\bm k})}{2T}
\right).
\eeq
The inverse of the coupling constant is associated with the parameter $\Delta_0/\varepsilon_{\rm F}$ through the gap equation at $T=B=0$, where $\Delta_0$ is the superfluid gap at $T=0$ and $B=0$. To quantify the finite-size correction of the neutron Fermi surface and particle-hole asymmetry, we introduce the parameter 
\beq
\delta \equiv \frac{N_{{\rm F},\uparrow}-N_{{\rm F},\downarrow}}{N_{{\rm F},\uparrow}+N_{{\rm F},\downarrow}}.
\label{eq:pha}
\eeq

To determine the critical temperatures, we linearize the gap equations \eqref{eq:gap1} and \eqref{eq:gap2} with respect to $\Delta$. The critical temperatures in the nonmagnetic $D_4$-BN state ($T_{{\rm c},0}$) and in the spin-polarized $^3P_2$ state ($T_{{\rm c},\kappa}$) are given by 
\beq
\frac{T_{{\rm c},\kappa}}{T_{{\rm c},0}} = \exp\left[
-\frac{3}{2{v}\bar{N}_{\rm F}}\left\{
\left( 1+ \frac{2\kappa}{1+\kappa^2}\delta\right)^{-1}-1
\right\}
\right],
\label{eq:tc_kappa}
\eeq
where $\bar{N}_{\rm F}\equiv ({N}_{{\rm F},\uparrow}+{N}_{{\rm F},\downarrow})/2$. 
In the spin-polarized state with $\kappa > 0$ ($\kappa< 0$), the spins of Cooper pairs are aligned parallel (anti-parallel) to the applied magnetic field. The finite-size correction of the neutron Fermi surface, $\delta$, splits the critical temperatures, which is crucial for the thermodynamic stability of the magnetized BN and ferromagnetic states with $\kappa > 0$ at high magnetic fields.
As shown in Fig.~\ref{fig:splitting}(b), the critical temperature of the magnetic state ($\kappa\neq 0$) is the same as that of the nonmagnetic $D_4$-BN state at the quasiclassical (particle-hole symmetric) limit, $\delta=0$, while $T_{{\rm c},\kappa}$ for $\kappa > 0$ increases as the particle-hole asymmetry parameter $\delta$ increases. This indicates that the spin-polarized states, such as the FM state and magnetized $D_4$-BN state, can be thermodynamically stable in a strong magnetic field, where the magnetic field gives rise to the Fermi level shift and particle-hole asymmetry of spin-up and down neutrons.

\subsection{$^3P_2$ superfluid phases under a magnetic field}

\begin{figure}[t!]
\includegraphics[width=85mm]{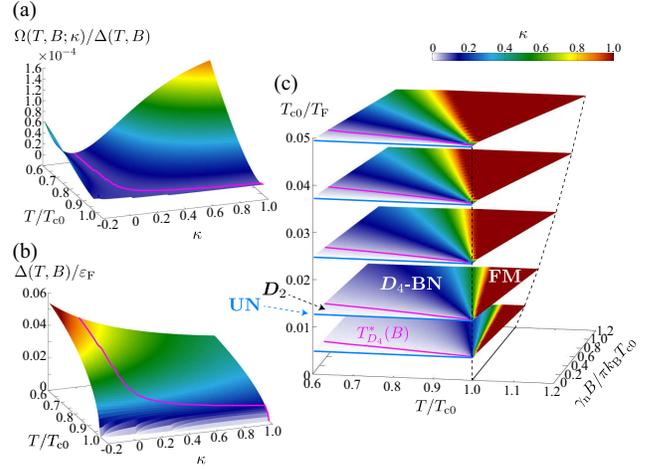}
\caption{Thermodynamic potential $\Omega(\kappa)$ (a) and the superfluid gap $\Delta(\kappa)$ (b) as functions of the temperature $T$ and the spin polarization of Cooper pairs, $\kappa$, at $\gamma _n \hbar B/\pi k_{\rm B}T_{\rm c0}=0.42$, where we set $T_{\rm c0}/T_{\rm F}=0.025$. In this high magnetic field regime, $r=-1$ is favored in the entire $T$. (c) $^3P_2$ superfluid phase diagram in the $T$-$B$ plane for various $T_{\rm c0}/T_{\rm F}$. In all data, we take $F^{({\rm s})}_0=F_0^{({\rm a})}=0$.}
\label{fig:data}
\end{figure}

Let us now clarify the thermodynamically stable $^3P_2$ superfluid phases under a strong magnetic field. In the previous works based on the quasiclassical approximation~\cite{Mizushima:2016fbn,Mizushima:2019spl}, we find that the UN phase appears at $B=0$, while the $D_2$ and $D_4$-BN states with $\kappa = 0$ are thermodynamically stable in the presence of magnetic fields. As mentioned in Sec.~\ref{sec:PHS}, the approximation postulates that the neutron Fermi surface is infinitely large, and the Zeeman splitting effect of the neutron Fermi surface becomes negligible. In the quasiclassical limit, $T_{\rm c0}/ T_{\rm F}\rightarrow 0$, therefore, a magnetic Zeeman field along the $\hat{z}$ direction gives rise to the pair breaking effect of the spin state, $\ket{\uparrow\downarrow+\downarrow\uparrow}$, while it does not affect the spin configurations of $\ket{\uparrow\uparrow}$ and $\ket{\downarrow\downarrow}$, leading to $\Delta_{\uparrow\uparrow}=\Delta_{\downarrow\downarrow}$. As the Cooper pairs in the $D_4$-BN state are equally distributed to the spin states, $\ket{\uparrow\uparrow}$ and $\ket{\downarrow\downarrow}$, it is tolerant against pair breaking effect by magnetic fields. The lower and higher region of the magnetic field in the phase diagram are occupied by the $D_2$ and $D_4$-BN phases, respectively, and the two phases are separated by the second- (first-) order phase boundary in the higher (lower) temperature regime. The phase boundaries meet at the (tri)critical end-point~\cite{Mizushima:2019spl}.

\begin{figure*}[t!]
\includegraphics[width=170mm]{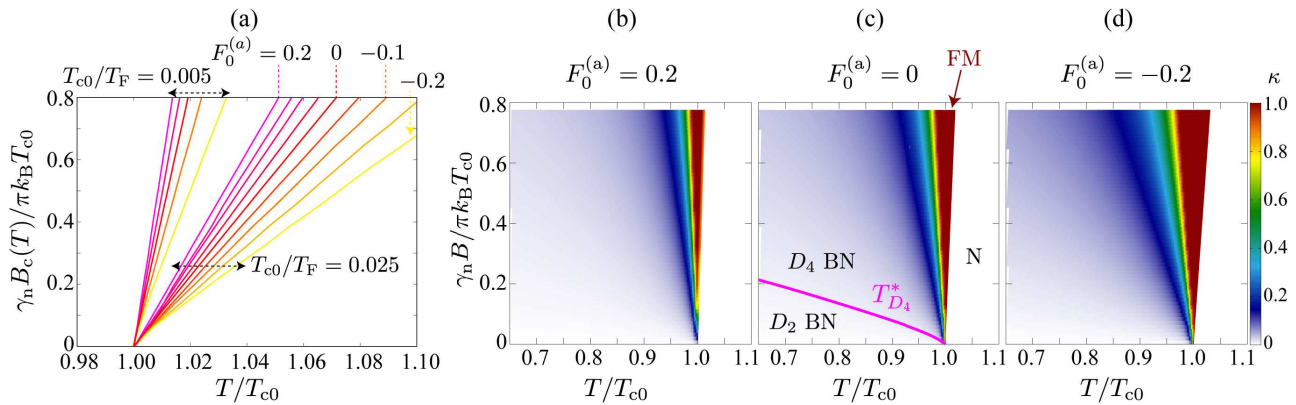}
\caption{(a) Critical magnetic fields $B_{\rm c}(T)$ in the fully spin-polarized $^3P_2$ state for several values of $F^{({\rm s})}_0$ (0.05 for each interval in $T_{\rm c0}/T_{\rm F}=0.025$ and 0.1 for each interval in $T_{\rm c0}/T_{\rm F}=0.005$), where $T_{\rm c0}$ is the critical temperature at zero magnetic fields. Phase diagram in the $T$-$B$ plane for $F^{({\rm a})}_0=0.2$ (b), $0$ (c), and $-0.2$ (d) for $T_{\rm c0}/T_{\rm F}=0.005$, where $T^{\ast}_{D_4}$ is the phase boundary between $D_2$- and $D_4$-BN states.}
\label{fig:tchf}
\end{figure*}

Here we determine the order parameters $\Delta(T,B)$, $r(T,B)$, and $\kappa(T,B)$ in Eq.~\eqref{eq:OP}, by numerically solving the self-consistent equations in Sec.~\ref{sec:scf} without employing the quasiclassical approximation. We introduce the dimensionless parameter 
\beq
T_{\rm c0}/T_{\rm F} \sim \Delta_0/\varepsilon_{\rm F}.
\eeq
This is related to the dimensionless parameter $\delta$ in Eq.~\eqref{eq:pha} representing the particle-hole asymmetry due to the finite size correction of the neutron Fermi surface. The quasiclassical limit corresponds to $T_{\rm c0}/T_{\rm F}\rightarrow 0$ and $\delta \rightarrow 0$, while the finite size correction of the Fermi surface is taken into account as $T_{\rm c0}/T_{\rm F}$ increases. In Figs.~\ref{fig:data}(a) and ~\ref{fig:data}(b), we plot the thermodynamic potential $\Omega (T,B)$ and the superfluid gap $\Delta(T,B)$, respectively, as functions of the nonunitarity $\kappa$, where we set $T_{\rm c0}/T_{\rm F}=0.025$ and $\gamma _n \hbar B/\pi k_{\rm B}T_{\rm c0}=0.42$. In this high magnetic field regime, $r=-1$ is favored in the entire $T$. The spin polarization of Cooper pairs, $\kappa$, stays zero in the low temperature region, which coincides with the superfluid phase diagram within the quasiclassical limit. As $T$ approaches $T_{\rm c0}$, however, $\kappa$ gradually increases and reaches $\kappa = 1$. The region with $\kappa \in (0,1)$ and $r=-1$ corresponds to the magnetized $D_4$-BN state with nonzero spin polarization, where Cooper pairs are condensed into $\uparrow\uparrow$ and $\downarrow\downarrow$ spin states but the population is imbalanced as shown in Eqs.~\eqref{eq:D41} and \eqref{eq:D42}.

The FM state with $\kappa = 1$ and $r=-1$ appears in the narrow region around $T_{\rm c}$. In Fig.~\ref{fig:data}(c), we plot the $^3P_2$ superfluid phase diagram in the $T$-$B$ plane for $0.005\le T_{\rm c0}/T_{\rm F} \le 0.05$. The stability region of the fully spin-polarized state enlarges and the critical temperature $T_{\rm c}$, corresponding to Eq.~\eqref{eq:tc_kappa} with $\kappa=1$, increases as $T_{\rm c0}/T_{\rm F}$ increases. In contrast to the enlargement of the spin-polarized state, the phase boundary between the $D_2$- and $D_4$-BN states, $T^{\ast}_{D_4}$, is insensitive to the parameter $T_{\rm c0}/T_{\rm F}$.

Let us now discuss the role of the Fermi liquid parameter, $F^{({\rm a})}_0$. The parameter $F^{({\rm a})}_0<0$ ($F^{({\rm a})}_0>0$) induces the Fermi liquid correction associated with a ferromagnetic (anti-ferromagnetic) exchange interaction, which alters the effective magnetic field via Eq.~\eqref{eq:mu_B}.
Here we ignore the contribution of other Fermi liquid parameters, e.g., $F^{({\rm s})}_0$, since it is renormalized to the chemical potential, and $\mu \approx \varepsilon_{\rm F}\gg F^{({\rm s})}_0n/N_{\rm F}$.
Figure \ref{fig:tchf}(a) shows the shift of the critical magnetic fields, $B_{\rm c}(T)$, in the FM state ($\kappa=1$ in Eq.~\eqref{eq:OP}) for several values of $F^{({\rm a})}_0$ (0.05 for each interval in $T_{\rm c0}/T_{\rm F}=0.025$ and 0.1 for each interval in $T_{\rm c0}/T_{\rm F}=0.005$). For nonzero $F^{({\rm a})}_0$, the local magnetization density induced by the spin-polarized $^3P_2$ pairing and the external magnetic field is fed back to the effective magnetic field in Eq.~\eqref{eq:mu_B}. As shown in Figs.~\ref{fig:tchf}(b-d), for $F^{({\rm a})}_0<0$ ($F^{({\rm a})}_0>0$), the induced ${\bm M}$ enhances (decreases) the effective magnetic field via the ferromagnetic (antiferromagnetic) spin exchange interaction. This effective magnetic field enlarges (suppresses) the thermodynamically stable region of the  FM state as $B$ increases. We note that at $B=0$, all the possible $^3P_2$ pairing state has the same critical temperature $T_{\rm c}=T_{\rm c0}$, regardless of $F^{({\rm s})}_0$. 


{In the high magnetic field regime, the neutron $^3P_2$ superfluid undergoes double second-order phase transitions at $T_{{\rm c},-}$ and $T_{{\rm c},+}$: the transition from the magnetized $D_4$-BN ($\kappa < 1$) to the FM state ($\kappa=1$) at $T_{{\rm c},-}$ and the transition from the FM state to the normal state at $T_{{\rm c},+}$. The signature of the multiple phase transitions is reflected as the successive jumps of the specific heat of neutrons, $C_{V} = - T\partial^2\Omega/\partial T^2$. It has been demonstrated in the previous work~\cite{Mizushima:2016fbn,Mizushima:2019spl} that the symmetry breaking driven by the phase transition from the $D_4$-BN to $D_2$-BN phases gives rise to the discontinuity of the specific heat jumps. Similarly, the multiple phase transitions are accompanied by the successive jumps of the specific heat at $T_{{\rm c},-}$ and $T_{{\rm c},+}$. The first jump is attributed to the condensation of only up-spin neutrons into $^3P_2$ Cooper pairs with orbital angular momentum $L_z=+1$, $\Delta_{\uparrow\uparrow}\propto (\hat{k}_x+i\hat{k}_y)$, while the second jump involves another spin component of neutrons to condense into the Cooper pairs with $L_z=-1$, $\Delta_{\downarrow\downarrow}\propto (\hat{k}_x-i\hat{k}_y)$.}


\section{Ginzburg-Landau theory for $^3P_2$ superfluids}
\label{sec:gl}

Around the transition point from the normal phase to the superfluid phase, the $^3P_2$ superfluids allow bosonic excitations as collective modes~\cite{Bedaque:2003wj,Leinson:2011wf,Leinson:2012pn,Leinson:2012pn,bedaquePRC14E,bedaquePRC14,Bedaque:2013fja,Bedaque:2014zta,Leinson:2009nu,Leinson:2010yf,Leinson:2010pk,Leinson:2010ru,Leinson:2011jr}.
Such modes are important not only in microscopic physics but also in astrophysics, because they can be relevant to the cooling process by neutrino emissions from neutron stars.\footnote{Note that the cooling process is relevant also to quantum vortices~\cite{Shahabasyan:2011zz}.}
Generally, bosonic excitations around the transition point can be described by the GL theory as a low-energy effective theory~\cite{fujitaPTP72,Richardson:1972xn,saulsPRD78,muzikarPRD80,saulsPRD82,Vulovic:1984kc,Masuda:2015jka,Chatterjee:2016gpm,Masuda:2016vak,Yasui:2018tcr,Yasui:2019tgc,Yasui:2019unp,Yasui:2019pgb,Yasui:2019vci,Mizushima:2019spl}.
The GL equation is expressed by a series of power terms of the order parameter
as a systematic expansion.
This is obtained through the fermion loops by integrating out the fermion degrees of freedom.

In many cases, the GL expansion may be given up to the fourth order. 
In the case of $^{3}P_{2}$ superfluids, however,
the expansion up to the fourth order is not enough to determine uniquely the ground state
due to a continuous degeneracy among 
the UN, $D_{2}$-BN, and $D_{4}$-BN phases at the fourth order.
At this order, in fact, an extended symmetry, SO(5), happens to exist thought it is absent in the original Hamiltonian.\footnote{In this case, the spontaneous breaking of the extended symmetry eventually generates a quasi-Nambu-Goldstone mode 
\cite{uchino}.
}
Such degeneracy can be resolved at the sixth order in the GL expansion~\cite{Masuda:2015jka}.
However, the sixth-order term is still not enough because it brings only the local stability for the small amplitude of the order parameter and it does not ensure the global stability for the large amplitude.
This problem can be cured at the eighth order~\cite{Yasui:2019unp}.
Therefore, the expansion up to the eighth order is
the minimum set of terms for allowing the ground state with uniqueness and globally stability. 
As a by-product, the expansion up to the eighth order enables us to identify the (tri)critical end-point separating the transition lines of 
the first- and second-order between 
the $D_2$-BN and $D_4$-BN phases~\cite{Mizushima:2019spl}.\footnote{The existence of the critical end-point was first found in the analysis of the BdG equation~\cite{Mizushima:2016fbn}.}

So far the GL expansion was given under the quasiclassical approximation at high-density region.
In this approximation, the coefficients of the GL equation can be expressed in simple forms according to the scaling behavior.
At low density, however, the quasiclassical approximation is not ensured anymore, and its violation should be considered carefully, as we have already discussed for the BdG equation in Sec.~\ref{sec:phase}. The violation of the quasiclassical approximation is relevant to the curvature of the Fermi surface, and it leads to the modification of the coefficients in the GL equation (see $\beta^{(1)}$ and $\gamma^{(1)}$ in Eqs.~\eqref{eq:gamma1_def} and \eqref{eq:beta1_def}).
After some calculations, we obtain the GL free energy expressed by
\begin{align}
  f(\tau)
=&
   f_{8}^{(0)}(\tau) + f_{2}^{(\le4)}(\tau) + f_{4}^{(\le2)}(\tau)  \nn \\
   & + {\cal O}(B^{m}\tau^{n})_{m+n\ge7},
\label{eq:eff_pot_coefficient02_f2}
\end{align}
where $\tau$ is a complex $3\times3$ matrix for the $^{3}P_{2}$ order parameter.
In the GL equation, we use the notation $\tau_{\mu i}$ for the order parameter of the $^{3}P_{2}$ superfluids, which is different only by a factor from $\mathcal{A}_{\mu i}$ used in the BdG theory in Sec.~\ref{sec:theory}.
Each term in Eq.~\eqref{eq:eff_pot_coefficient02_f2} is explained as the followings.
$f_{8}^{(0)}(\tau)$ is the term up to the eighth order of the superfluid condensate without magnetic fields:
\begin{widetext}
\begin{eqnarray}
 f_{8}^{(0)}(\tau)
&=&
  K^{(0)}
  \Bigl(
        \nabla_{xi} \tau_{\mu j \ast}
        \nabla_{xi} \tau_{\mu j}
     + \nabla_{xi} \tau_{\mu i\ast}
        \nabla_{xj} \tau_{\mu j}
     + \nabla_{xi} \tau_{\mu j \ast}
        \nabla_{xj} \tau_{\mu i}
  \Bigr)
\nonumber \\ &&
+ \alpha^{(0)}
   \mathrm{tr}\bigl( \tau^{\ast} \tau \bigr)
\nonumber \\ &&
+ \beta^{(0)}
   \Bigl(
        \mathrm{tr}\bigl( \tau^{\ast} \tau \bigr) \mathrm{tr}\bigl( \tau^{\ast} \tau \bigr)
      - \mathrm{tr}\bigl( \tau^{\ast} \tau^{\ast} \tau \tau \bigr)
   \Bigr)
\nonumber \\ &&
+ \gamma^{(0)}
   \Bigl(
         - 3 \mathrm{tr}\bigl( \tau \tau^{\ast} \bigr) \, \mathrm{tr}\bigl( \tau \tau \bigr) \, \mathrm{tr}\bigl( \tau^{\ast} \tau^{\ast} \bigr)
        + 4 \mathrm{tr}\bigl( \tau \tau^{\ast} \bigr) \, \mathrm{tr}\bigl( \tau \tau^{\ast} \bigr) \, \mathrm{tr}\bigl( \tau \tau^{\ast} \bigr)
              \nonumber \\ && \hspace{3em} 
        + 6 \mathrm{tr}\bigl( \tau^{\ast} \tau \bigr) \, \mathrm{tr}\bigl( \tau^{\ast} \tau^{\ast} \tau \tau \bigr)
      + 12 \mathrm{tr}\bigl( \tau^{\ast} \tau \bigr) \, \mathrm{tr}\bigl( \tau^{\ast} \tau \tau^{\ast} \tau \bigr)
         - 6 \mathrm{tr}\bigl( \tau^{\ast} \tau^{\ast} \bigr) \, \mathrm{tr}\bigl( \tau^{\ast} \tau \tau \tau \bigr)
         - 6 \mathrm{tr}\bigl( \tau \tau \bigr) \, \mathrm{tr}\bigl( \tau^{\ast} \tau^{\ast} \tau^{\ast} \tau \bigr)
              \nonumber \\ && \hspace{3em} 
       - 12 \mathrm{tr}\bigl( \tau^{\ast} \tau^{\ast} \tau^{\ast} \tau \tau \tau \bigr)
      + 12 \mathrm{tr} \bigl( \tau^{\ast} \tau^{\ast} \tau \tau \tau^{\ast} \tau \bigr)
        + 8 \mathrm{tr}\bigl( \tau^{\ast} \tau \tau^{\ast} \tau \tau^{\ast} \tau \bigr)
   \Bigr)
\nonumber \\ &&
 + \delta^{(0)}
\Bigl(
       \bigl( \mathrm{tr}\,\tau^{\ast 2} \bigr)^{2} \bigl( \mathrm{tr}\, \tau^{2} \bigr)^{2}
 + 2 \bigl( \mathrm{tr}\,\tau^{\ast 2} \bigr)^{2} \bigl( \mathrm{tr}\, \tau^{4} \bigr)
  - 8 \bigl( \mathrm{tr}\,\tau^{\ast 2} \bigr)
       \bigl( \mathrm{tr}\,\tau^{\ast}\tau\tau^{\ast}\tau \bigr)
       \bigl( \mathrm{tr}\,\tau^{2} \bigr)
  - 8 \bigl( \mathrm{tr}\,\tau^{\ast 2} \bigr) \bigl( \mathrm{tr}\,\tau^{\ast}\tau \bigr)^{2} \bigl( \mathrm{tr}\,\tau^{2} \bigr)
       \nonumber \\ && \hspace{3em}
 - 32 \bigl( \mathrm{tr}\,\tau^{\ast 2} \bigr) \bigl( \mathrm{tr}\,\tau^{\ast}\tau \bigr) \bigl( \mathrm{tr}\,\tau^{\ast}\tau^{3} \bigr)
 - 32 \bigl( \mathrm{tr}\,\tau^{\ast 2} \bigr) \bigl( \mathrm{tr}\,\tau^{\ast}\tau\tau^{\ast}\tau^{3} \bigr)
 - 16 \bigl( \mathrm{tr}\,\tau^{\ast 2} \bigr) \bigl( \mathrm{tr}\,\tau^{\ast}\tau^{2}\tau^{\ast}\tau^{2} \bigr)
       \nonumber \\ && \hspace{3em}
  + 2 \bigl( \mathrm{tr}\,\tau^{\ast 4} \bigr) \bigl( \mathrm{tr}\,\tau^{2} \bigr)^{2}
  + 4 \bigl( \mathrm{tr}\,\tau^{\ast 4} \bigr) \bigl( \mathrm{tr}\,\tau^{4} \bigr)
  - 32 \bigl( \mathrm{tr}\,\tau^{\ast 3}\tau \bigr) \bigl( \mathrm{tr}\,\tau^{\ast}\tau \bigr) \bigl( \mathrm{tr}\,\tau^{2} \bigr)
       \nonumber \\ && \hspace{3em}
  - 64 \bigl( \mathrm{tr}\,\tau^{\ast 3}\tau \bigr) \bigl( \mathrm{tr}\,\tau^{\ast}\tau^{3} \bigr)
  - 32 \bigl( \mathrm{tr}\,\tau^{\ast 3}\tau\tau^{\ast}\tau \bigr) \bigl( \mathrm{tr}\,\tau^{2} \bigr)
  - 64 \bigl( \mathrm{tr}\,\tau^{\ast 3}\tau^{2}\tau^{\ast}\tau^{2} \bigr)
  - 64 \bigl( \mathrm{tr}\,\tau^{\ast 3}\tau^{3} \bigr) \bigl( \mathrm{tr}\,\tau^{\ast}\tau \bigr)
       \nonumber \\ && \hspace{3em}
  - 64 \bigl( \mathrm{tr}\,\tau^{\ast 2}\tau\tau^{\ast 2}\tau^{3} \bigr)
  - 64 \bigl( \mathrm{tr}\,\tau^{\ast 2}\tau\tau^{\ast}\tau^{2} \bigr) \bigl( \mathrm{tr}\,\tau^{\ast}\tau \bigr)
 + 16 \bigl( \mathrm{tr}\,\tau^{\ast 2}\tau^{2} \bigr)^{2}
 + 32 \bigl( \mathrm{tr}\,\tau^{\ast 2}\tau^{2} \bigr) \bigl( \mathrm{tr}\,\tau^{\ast}\tau \bigr)^{2}
       \nonumber \\ && \hspace{3em}
 + 32 \bigl( \mathrm{tr}\,\tau^{\ast 2}\tau^{2} \bigr) \bigl( \mathrm{tr}\,\tau^{\ast}\tau\tau^{\ast}\tau \bigr)
 + 64 \bigl( \mathrm{tr}\,\tau^{\ast 2}\tau^{2}\tau^{\ast 2}\tau^{2} \bigr)
  -16 \bigl( \mathrm{tr}\,\tau^{\ast 2}\tau\tau^{\ast 2}\tau \bigr) \bigl( \mathrm{tr}\,\tau^{2} \bigr)
 + 8 \bigl( \mathrm{tr}\,\tau^{\ast}\tau \bigr)^{4}
       \nonumber \\ && \hspace{3em}
 + 48 \bigl( \mathrm{tr}\,\tau^{\ast}\tau \bigr)^{2} \bigl( \mathrm{tr}\,\tau^{\ast}\tau\tau^{\ast}\tau \bigr)
 +192 \bigl( \mathrm{tr}\,\tau^{\ast}\tau \bigr) \bigl( \mathrm{tr}\,\tau^{\ast}\tau\tau^{\ast 2}\tau^{2} \bigr)
 + 64 \bigl( \mathrm{tr}\,\tau^{\ast}\tau \bigr) \bigl( \mathrm{tr}\,\tau^{\ast}\tau\tau^{\ast}\tau\tau^{\ast}\tau \bigr)
       \nonumber \\ && \hspace{3em}
  -128 \bigl( \mathrm{tr}\,\tau^{\ast}\tau\tau^{\ast 3}\tau^{3} \bigr)
 + 64 \bigl( \mathrm{tr}\,\tau^{\ast}\tau\tau^{\ast 2}\tau\tau^{\ast}\tau^{2} \bigr)
 + 24 \bigl( \mathrm{tr}\,\tau^{\ast}\tau\tau^{\ast}\tau \bigr)^{2}
 +128 \bigl( \mathrm{tr}\,\tau^{\ast}\tau\tau^{\ast}\tau\tau^{\ast 2}\tau^{2} \bigr)
       \nonumber \\ && \hspace{3em}
 + 48 \bigl( \mathrm{tr}\,\tau^{\ast}\tau\tau^{\ast}\tau\tau^{\ast}\tau\tau^{\ast}\tau \bigr)
\Bigr).
\end{eqnarray}
\end{widetext}
Notice that the $\delta^{(0)}$ term is the eighth-order term for the global stability which was found in Ref.~\cite{Yasui:2019unp}.
The other two terms in Eq.~\eqref{eq:eff_pot_coefficient02_f2} contain the magnetic fields:
\begin{eqnarray}
   f_{2}^{(\le4)}(\tau)
&=&
         \beta^{(1)}
	 i B^{k} \,
 	 \mathrm{tr}\bigl(\epsilon^{k}\tau \tau^{\ast}\bigr)
	+ \beta^{(2)}
	   \vec{B}^{t} \tau \tau^{\ast} \vec{B}
	+ \beta^{(4)}
	        |\vec{B}|^{2}
	        \vec{B}^{t} \tau \tau^{\ast} \vec{B},
\nonumber \\
\end{eqnarray}
and
\begin{eqnarray}
f_{4}^{(\le2)}(\tau)
&&=
	   \gamma^{(1)}
	   i B^{k}
	   \Bigl(
	        \mathrm{tr}\bigl( \epsilon^{k} \tau \tau^{\ast} \bigr)
	        \mathrm{tr}\bigl( \tau \tau^{\ast} \bigr)\nn\\
	        && \hspace{4em}
	      - \mathrm{tr}\bigl( \epsilon^{k} \tau \tau \tau^{\ast} \tau^{\ast} \bigr)
	      - \mathrm{tr}\bigl( \epsilon^{k} \tau \tau^{\ast} \tau \tau^{\ast} \bigr)
	  \Bigr)
	  \nonumber \\ && 
	+
	  \gamma^{(2)}
	  \Bigl(
	          - 2 \, |\vec{B}|^{2} \, \mathrm{tr}\bigl( \tau \tau \bigr) \, \mathrm{tr}\bigl( \tau^{\ast} \tau^{\ast} \bigr)
	          - 4 \, |\vec{B}|^{2} \, \mathrm{tr}\bigl( \tau \tau^{\ast} \bigr) \, \mathrm{tr}\bigl( \tau \tau^{\ast} \bigr)
	             \nonumber \\ && \hspace{2em}
	         + 4 \, |\vec{B}|^{2} \, \mathrm{tr}\bigl( \tau \tau^{\ast} \tau \tau^{\ast} \bigr)
	         + 8 \, |\vec{B}|^{2} \, \mathrm{tr}\bigl( \tau \tau \tau^{\ast} \tau^{\ast} \bigr)
	             \nonumber \\ && \hspace{2em}
	         + \vec{B}^{t} \tau \tau \vec{B} \, \mathrm{tr}\bigl( \tau^{\ast} \tau^{\ast} \bigr)
	         + \vec{B}^{t} \tau^{\ast} \tau^{\ast} \vec{B} \, \mathrm{tr}\bigl( \tau \tau \bigr)
	             \nonumber \\ && \hspace{2em}
	          - 8 \, \vec{B}^{t} \tau \tau^{\ast} \vec{B} \, \mathrm{tr}\bigl( \tau \tau^{\ast} \bigr)
	         + 2 \, \vec{B}^{t} \tau \tau^{\ast} \tau^{\ast} \tau \vec{B}
	             \nonumber \\ && \hspace{2em}
	         + 2 \, \vec{B}^{t} \tau^{\ast} \tau \tau \tau^{\ast} \vec{B}
	          - 8 \, \vec{B}^{t} \tau \tau \tau^{\ast} \tau^{\ast} \vec{B}
	             \nonumber \\ && \hspace{2em}
	          - 8 \, \vec{B}^{t} \tau \tau^{\ast} \tau \tau^{\ast} \vec{B}
	  \Bigr).
\end{eqnarray}
In the above equations, the coefficients are uniquely determined by the loop calculation for fermion,
\begin{eqnarray}
   K^{(0)} &\equiv& \frac{7\zeta(3)N_{0}k_{\rm F}^{4}}{240m^{2}(\pi T_{\rm c0})^{2}},
\\ 
   \alpha^{(0)} &\equiv& \frac{N_{0}k_{\rm F}^{2}}{3} \frac{T-T_{\rm c0}}{T_{\rm c0}},
\\ 
   \beta^{(0)} &\equiv& \frac{7\zeta(3)N_{0}k_{\rm F}^{4}}{60(\pi T_{\rm c0})^{2}},
\\ 
   \gamma^{(0)} &\equiv& -\frac{31\zeta(5)N_{0}k_{\rm F}^{6}}{13440(\pi T_{\rm c0})^{4}},
\\ 
   \delta^{(0)} &\equiv& \frac{127\zeta(7)N_{0}k_{\rm F}^{8}}{387072(\pi T_{\rm c0})^{6}},
\\ 
   \beta^{(2)} &\equiv& \frac{7\zeta(3)N_{0}k_{\rm F}^{2}}{12(\pi T_{\rm c0})^{2}} |\vec{m}_{n}^{\ast}|^{2},
\\ 
   \beta^{(4)} &\equiv& -\frac{31\zeta(5)N_{0}k_{\rm F}^{2}}{48(\pi T_{\rm c0})^{4}} |\vec{m}_{n}^{\ast}|^{4},
\\ 
   \gamma^{(2)} &\equiv& \frac{31\zeta(5)N_{0}k_{\rm F}^{4}}{960(\pi T_{\rm c0})^{4}} |\vec{m}_{n}^{\ast}|^{2},
\end{eqnarray}
and
\begin{eqnarray}
   \gamma^{(1)} &\equiv& \frac{7\zeta(3)N_{0}m\,k_{\rm F}^{2}}{24(\pi T_{\rm c0})^{2}} |\vec{m}_{n}^{\ast}|,
\label{eq:gamma1_def} \\ 
   \beta^{(1)} &\equiv& -\frac{N_{0}m}{2} J_{1}(D/T_{\rm c0}) |\vec{m}_{n}^{\ast}|,
\label{eq:beta1_def}
\end{eqnarray}
with the state-number density $N_{0}=m\,k_{\rm F}/(2\pi^{2})$ at the Fermi surface, the magnitude of the magnetic moment $|\vec{m}_{n}^{\ast}|=(\gamma_{n}/2)/(1+F_{0}^{({\rm a})})$, the superfluid transition temperature at zero magnetic fields $T_{\rm c0}$, and the polarization vector $\epsilon^{k}$ ($k=1,2,3$) defined by
\begin{gather}
   \epsilon^{1}
=
   \left(
   \begin{array}{ccc}
    0 & 0  & 0  \\
    0 & 0  & 1  \\
    0 & -1 & 0  
   \end{array}
   \right),~
   \epsilon^{2}
=
   \left(
   \begin{array}{ccc}
    0 & 0  & -1  \\
    0 & 0  & 0  \\
    1 & 0 & 0  
   \end{array}
   \right), \nn \\
   \epsilon^{3}
=
   \left(
   \begin{array}{ccc}
    0 & 1  & 0  \\
    -1 & 0  & 0  \\
    0 & 0 & 0  
   \end{array}
   \right).
\label{eq:epsilon_matrix_def2}
\end{gather}
We note that $K^{(0)}$, $\alpha^{(0)}$, $\beta^{(0)}$, $\gamma^{(0)}$, $\delta^{(0)}$, $\beta^{(2)}$ and $\beta^{(4)}$ are obtained under the qausiclassical approximation, where the Fermi surface is regarded as the flat surface approximately.
On the other hand, the $\beta^{(1)}$ and $\gamma^{(1)}$ terms represent the violation of the particle-hole symmetry.
Those two terms are beyond the quasiclassical approximation and they are relevant to the curvature of the Fermi surface.
In the definition of $\beta^{(1)}$, the function $J_{1}(x)$ is given by
\begin{eqnarray}
   J_{1}(D/T_{\rm c0})
&\equiv&
   \int_{-D/T_{\rm c0}}^{D/T_{\rm c0}}
   \frac
   {\bigl( \mathrm{sinh}\,x - x \bigr)\,\mathrm{sech}^{2}
   (x/2)}
   {2x}
   \mathrm{d}x > 0,
\label{eq:J1_definition}
\end{eqnarray}
where $D>0$ is the cutoff parameter for the energy measured from the Fermi surface: $\xi_{\vec{p}} \in [-D,D]$.
$D$ should be a finite value, because the scaling behavior does not hold at the lowest-order for the violation of the particle-hole symmetry.
In fact, the value of $J_{1}(D/T_{\rm c0})$ becomes divergent for the infinite value of $D$.
There is no such divergence in the coefficient $\gamma^{(1)}$ due to the higher-order loops of fermions.
We confirm that the $\beta^{(1)}$ and $\gamma^{(1)}$ terms vanish in the quasiclassical approximation with the large $k_{\rm F}$ limit.
Thus, those terms are relevant at the low density region only.

Here, we comment that the $\beta^{(1)}$ term was first given for the $^{3}\mathrm{He}$ superfluids in the GL formalism~\cite{AmbegaokarMermin1973}.
The authors in Ref.~\cite{AmbegaokarMermin1973}, however, regarded it as the phenomenological parameter whose values should be fitted to the experimental data.
In our work, in contrast, the $\beta^{(1)}$ term as well as the $\gamma^{(1)}$ term are determined microscopically in the expansions of the fermion loops.

Similarly to Eq.~\eqref{eq:OP}, we parametrize $\tau_{\mu i}$ by
\begin{eqnarray}
   \tau_{\mu i}
=
   \tau_{0}
   \left(
   \begin{array}{ccc}
    r & i\kappa & 0 \\
    i\kappa & 1 & 0 \\
    0 & 0 & -1-r 
   \end{array}
   \right)_{\mu i},
\label{eq:OP_NF_def}
\end{eqnarray}
with real parameters $\tau_{0}\ge0$, $-1\le r \le -1/2$, and $-1\le \kappa \le 1$.
We remind us that the amplitude $\tau_{0}$ can be different from that in the BdG theory due to the different overall factor as discussed already.
The nonzero value of $\kappa$ is relevant to the magnetized $D_4$-BN and FM phases.
The values of $\tau_{0}$, $r$, and $\kappa$ are determined uniquely by minimizing the GL free energy for given temperature, magnetic field and particle density.

\begin{figure}[tb]
\includegraphics[width=85mm]{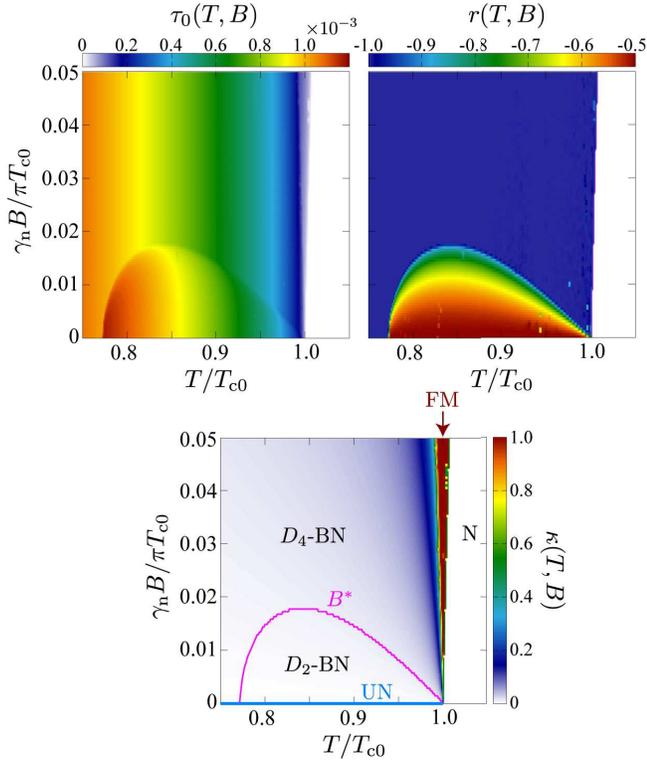}
\caption{The phase diagram by the GL equation: $\tau_{0}$, $r$, and $\kappa$ on the $T$-$B$ plane, where ``UN'', ``BN'', ``FM'', and ``N'' denote the uniaxial nematic, biaxial nematic, ferromagnetic, and normal states, respectively. We also plot the critical field between $D_2$- and $D_4$-BN states, $B^{\ast}$.}
\label{fig:Fig_190323_r}
\end{figure}

Let us show the numerical result of the phase diagram by the GL free energy. 
As the input parameter, we determine the cutoff parameter $D$ in the following way.
According to Ref.~\cite{takatsukaPTP71}, the momentum cutoff parameter $\Lambda=1.509$ fm$^{-1}$ was introduced for the interaction potential between two neutrons.\footnote{Notice that $a_{\Lambda}$ is used as the momentum cutoff parameter in Ref.~\cite{takatsukaPTP71}.}
Utilizing this value as a reference, we estimate the value of $D$ as $D\approx\Lambda^{2}/(2m)=47.17$ MeV where $m$ is the mass of a neutron.
We show the result of $\tau_{0}$, $r$, and $\kappa$ on the $T$-$B$ plane in Fig.~\ref{fig:Fig_190323_r}.
It is confirmed that, in the superfluid phase, there exist the UN phase at zero magnetic field, the $D_2$-BN phase at weak magnetic fields, and the $D_4$-BN phase at strong magnetic fields, as known in the past studies.
Interestingly, we find that that the FM phase ($\kappa =1$) appears around the boundary region between the $D_4$-BN phase and the normal phase at nonzero magnetic fields.
This is consistent with the result in the self-consistent calculation of the thermodynamic potential for fermions (Fig.~\ref{fig:data}).
In the GL equation, the appearance of the FM phase is caused mainly by the $\beta^{(1)}$ term.\footnote{Because the FM fields exist at nonzero magnetic fields, it is enough to consider only the terms coupling to the magnetic fields, such as the $\beta^{(1)}$ term, as relevant terms in the next-to-leading order.}
Therefore, the $\beta^{(1)}$ term in the GL equation is essentially important for realizing the spin-polarized phase beyond the quasiclassical approximation.

\section{Concluding remarks}
\label{sec:summary} 

We have investigated the thermodynamic stability of neutron $^3P_2$ superfluids under a strong magnetic field. A $^3P_2$ superfluid is expected to exist in the interior of neutron stars under extreme conditions. In particular, magnetars have the most intense magnetic field in the Universe. The strong field, $B = 10^{15}$-$10^{18}~{\rm G}$, amounts the critical field ($\sim k_{\rm B}T_{\rm c}$) of spin-singlet Cooper pairing, corresponding to $\gamma \hbar B /\pi k_{\rm B}T_{\rm c}\sim 0.1$-$100$ for $T_{\rm c}=10^9~{\rm K}$. The interaction of neutron spins with such strong magnetic field breaks spin-singlet Cooper pairs, and the conventional $^1S_0$ superfluids are thermodynamically unstable against the Zeeman splitting of the neutron Fermi surface. Employing the self-consistent calculations properly incorporating the Zeeman effect, we have demonstrated that spin-polarized $^3P_2$ Cooper pairs are tolerant against a strong magnetic field. A Zeeman field induces the successive phase transitions from the UN state ($r=-1/2$ and $\kappa=0$) to the $D_2$- and $D_4$-BN states ($r\in(-1,1/2)$ and $\kappa\in (0,1)$) and the FM state ($r=-1$ and $\kappa = 1$). The variational parameters, $r\in [-1,-1/2]$ and $\kappa\in[0,1]$, quantify the biaxility and the spin polarization of $^3P_2$ order parameters, respectively. The FM state becomes thermodynamically stable in the high temperature and strong magnetic field, and the critical temperature rather increases with increasing $B$. Therefore, the spin-polarized $^3P_2$ superfluid is expected to survive in the interior of magnetars with intense magnetic field.

Neutron $^3P_2$ superfluids have two key ingredients, Bogoliubov quasiparticles and $^3P_2$ Cooper pairs. The former is the elementary fermionic excitations from the superfluid ground state, while the latter behaves as long-lived bosons with internal degrees of freedom $J_z= 0, \pm 1, \pm 2$. The superfluids share much common interests in condensed-matter physics, including $d$-wave superconductors~\cite{Mermin:1974zz}, the superfluid $^3$He~\cite{vollhardt2013superfluid,volovik,mizushimaJPCM15,mizushimaJPSJ16}, and Weyl superconductivity, e.g., in heavy-fermion superconductors. All $^3P_2$ superfluid phases are prototypes of topological superconductors and Weyl superconductors, where low-lying Bogoliubov quasiparticles behave as Majorana and Weyl fermions~\cite{Mizushima:2016fbn}. In addition, the order parameter manifold is similar to that in spin-2 Bose-Einstein condensates~\cite{kawaguchi}. {The order parameter manifolds of magnetic $^3P_2$ superfluid phases, such as the magnetic $D_4$-BN and FM phases, are similar with those of the broken axisymmetric and FM phases in the context of spin-2 Bose-Einstein condensates, respectively, which bring about a variety of exotic topological defects~\cite{kobayashi21,kawaguchi}. Hence, neutron stars are colossal topological superfluidity in the Universe.}

The Bogoliubov quasiparticles can be main carriers in thermal and spin transport phenomena. In $^3P_2$ superfluids, their gap symmetries and topological properties are sensitive to the biaxility $r$ and nonunitarity $\kappa$. In the UN state, the excitation energy in Eq.~\eqref{eq:EQP} is gapped out on the entire Fermi surface of neutrons, and the excitation gap is uniaxially elongated in the momentum space. The uniaxial gap structure continuously deforms into $D_2$ and $D_4$ symmetric shape with changing the biaxiality $r$, and becomes gapless, $E({\bm k}_0)=0$, at the pairwise points, ${\bm k}_0= \pm k_{\rm F}\hat{\bm z}$, in the $D_4$-BN state, where the $z$-axis is parallel to the magnetic field. In the lower field, the BN state with $\kappa=0$ is stabilized, while the strong magnetic field favors the magnetized BN state with $\kappa\neq 0$. The gapless quasiparticle excitations in the case of $\kappa=0$ are spin degenerate, but the nonzero $\kappa$ induces the spin polarization of the superfluid ground states and thus the gapless excitations are magnetized. As $B$ further increases,  its value reaches $\kappa=1$ and the quasiparticle excitations are fully spin-polarized in the sense that only the $\uparrow\uparrow$ Cooper pairs survive and the other spin channels are broken by the strong Zeeman effect. The gapless points are protected by topological invariants and the low-lying Bogoliubov quasiparticles behave as Dirac and Weyl fermions. As shown in the current work, a strong magnetic field in neutron stars and magnetars gives rise to the successive phase transitions in $^3P_2$ superfluids and makes the quasiparticle spectrum anisotropic and magnetized. Hence, it is expected that thermal transport mediated by anisotropic and gapless Bogoliubov quasiparticles leads to the anisotropic thermal evolution of neutron stars. 
The thermal transport by Weyl and Dirac fermions in $^3P_2$ superfluid ground states remains as an important key for understanding the thermal evolution of neutron stars and magnetars. 


In addition to fermions, the condensation of $^3P_2$ Cooper pairs is accompanied by rich order parameter manifold and thus paves a way for understanding the thermal evolution of neutron stars and a longstanding issue on glitches in pulsars. 
As discussed in Sec.~\ref{sec:symmetry}, there are 
several Nambu-Goldstone bosons associated with nontrivial broken symmetries in $^3P_2$ ordered states. Such bosonic excitations might be relevant for the cooling process by neutrino emission~\cite{Bedaque:2003wj,Leinson:2011wf,Leinson:2012pn,Leinson:2013si,Bedaque:2012bs,bedaquePRC14,Bedaque:2013fja,Bedaque:2014zta,Leinson:2009nu,Leinson:2010yf,Leinson:2010pk,Leinson:2010ru,Leinson:2011jr}. 
The GL equation can be easily applied to describe the condensations nonuniform in space.
In $^{3}P_{2}$ superfluids, the GL equation was utilized to investigate  topological objects such as
spontaneously magnetized vortices~\cite{muzikarPRD80,saulsPRD82,fujitaPTP72,Masuda:2015jka,Masuda:2016vak},
vortices with Majorana fermions~\cite{Masaki:2019rsz}, 
solitonic excitations on a vortex~\cite{Chatterjee:2016gpm}, {coreless vortices~\cite{leinson20}}, non-Abelian half-quantum vortices~\cite{Masuda:2016vak,masaki}, and
topological defects (boojums) on the boundary of $^3P_2$ superfluids~\cite{Yasui:2019pgb}. The boojums on the boundary of $^3P_2$ superfluids have similarities to the objects on the boundary of spin-2 Bose-Einstein condensations~\cite{2019arXiv190702216C} and liquid crystals~\cite{Urbanski_2017}.
The GL equation was adopted to investigate the quasistable domain walls in the neutron $^{3}P_{2}$ superfluids~\cite{Yasui:2019vci}.
At lower density, the neutron $^1S_0$ superfluid plays the important role. Around the boundary region of the $^1S_0$ and $^3P_2$ superfluids, both phases can coexist as shown by the GL equation~\cite{Yasui:2020xqb}.
The effect of the spin-polarized $^3P_2$ phase on the $^1S_0$ superfluids is an interesting subject left for future works.

\begin{acknowledgments} 

We thank Michikazu Kobayashi for useful discussion. T.M. is supported by a Grant-in-Aid for Scientific Research on Innovative Areas ``Quantum Liquid Crystals (JP20H05163)'' from the Japan Society for the Promotion of Science (JSPS), and JSPS KAKENHI (Grant No.~JP20K03860, No.~JP20H01857, and No.~JP21H01039). The work of S.Y., D.I. and M.N. was supported by the Ministry of Education, Culture, Sports, Science (MEXT)-Supported Program for the Strategic Research Foundation at Private Universities Topological Science (Grant No. S1511006). S.Y. is supported by JSPS KAKENHI (Grant No. JP17K05435) and the Interdisciplinary Theoretical and Mathematical Sciences Program (iTHEMS) at RIKEN. D.I. is supported by the Financial Support of Fujukai Foundndation, and M.N. is supported in part by JSPS KAKENHI (Grant No.~JP18H01217).

\end{acknowledgments}


\bibliography{neutronstar}

\end{document}